\documentstyle[12pt]{article}
\newcommand{\pr}{\paragraph{}}
\newcommand{\be}{\begin{equation}}
\newcommand{\ee}{\end{equation}}
\def\lsim{\mathrel{\rlap {\raise.5ex\hbox{$ < $}}
{\lower.5ex\hbox{$\sim$}}}}

\newcommand{\bea}{\begin{eqnarray}}
\newcommand{\nn}{\nonumber}
\newcommand{\eea}{\end{eqnarray}}
\newcommand{\nd}[1]{/\hspace{-0.6em} #1}
\newcommand{\nk}{\noindent}
\begin{document}
\begin{titlepage}
\vspace{.015in}
\begin{flushright}
CERN-TH.7195/94 \\
ENS-LAPP-A-463/94 \\
ACT-5/94 \\
CTP-TAMU-13/94  \\
\end{flushright}

\begin{centering}

\vspace{.05 in}
{\Large {\bf A Non-Critical String Approach to
Black Holes, Time and Quantum Dynamics  }} \\
\vspace{.100in}
{\bf John Ellis$^{a,\diamond} $}, {\bf N.E. Mavromatos$^{b,\diamond }$}
and {\bf D.V.
Nanopoulos}$^{a,c}$   \\
\vspace{.45in}
\vspace{.05in}
{\bf Abstract} \\

\end{centering}

\paragraph{}
{\small We review our approach to time and quantum dynamics based on
non-critical string theory, developing its relationship to
previous work on non-equilibrium quantum statistical mechanics
and the microscopic arrow of time.
We exhibit specific non-factorizing contributions to the
${\nd S}$ matrix associated with topological defects on the
world sheet, explaining the r\^ole that the leakage of
${W_{\infty}}$ charges plays in the loss of quantum coherence.
We stress the analogy with the quantum Hall effect,
discuss the violation of $CPT$,
and
also apply our approach to cosmology.}

\par
\vspace{0.5in}

\vspace{.1in}

\begin{flushleft}
CERN-TH.7195/94 \\
ENS-LAPP-A-463/94 \\
ACT-5/94 \\
CTP-TAMU-13/94  \\
March 1994 (revised: July 1994) \\
\end{flushleft}
\vspace{.1in}

--------------------------------------------- \\
$^{a}$ Theory Division, CERN, CH-1211, Geneva 23, Switzerland  \\
$^{b}$ Laboratoire de Physique Th\`eorique
ENSLAPP (URA 14-36 du CNRS, associe\`e \`a l' E.N.S
de Lyon, et au LAPP (IN2P3-CNRS) d'Annecy-le-Vieux),
Chemin de Bellevue, BP 110, F-74941 Annecy-le-Vieux
Cedex, France. \\
On leave from S.E.R.C. Advanced Fellowship, Dept. of Physics
(Theoretical Physics), University of Oxford, 1 Keble Road,
Oxford OX1 3NP, U.K.  \\
$^{c}$ Center for
Theoretical Physics, Dept. of Physics,
Texas A \& M University, College Station, TX 77843-4242, USA
and Astroparticle Physics Group, Houston
Advanced Research Center (HARC), The Mitchell Campus,
Woodlands, TX 77381, USA. \\

\begin{centering}

$^\diamond $ {\bf Lectures presented at : \\
the Erice Summer School,
31st Course: From Superstrings to the Origin of
Space-Time, Ettore Majorana Centre, Erice,
July 4-12 1993 }\\

\end{centering}

\end{titlepage}

\newpage
\section{Introduction}
\pr
String theory is widely heralded as a consistent quantum theory
of gravity. As such, it should not only enable us to calculate
meaningfully quantum-gravitational corrections to scattering
processes in a fixed space-time background, but also to take into
account quantum fluctuations in space and time themselves. The way
to carry out the first part of this double programme in string
theory is well known: calculate higher-genus effects in a given
critical string vacuum that describes the appropriate classical
background. The way to carry out the second part of the quantum
gravity programme in string theory, namely
to understand space-time foam,
is less evident. One must
master a multitude of string vacua and the quantum transitions
between them, which necessarily involve non-critical string
theory \cite{aben3}.
No-one can do this at present: the best one can do is study
some tractable examples of non-critical string models \cite{emnqm}
that one
hopes are relevant and representative, and abstract from them
features that may be generic.
\pr
It is to be expected that some shibboleths of conventional physics
will be cast down when this programme is carried out. Certainly
general relativity must be modified, and perhaps also special
relativity and even quantum mechanics. We review in these lectures
our
work \cite{emnqm,emnshort,emndollar}
indicating that our understandings of special relativity,
quantum mechanics and quantum field theory should indeed be
modified. We start from the string black hole solution of
Witten \cite{witt},
and abstract from it general features associated with a
renormalization group analysis of the string effective
action \cite{rg}.
\pr
There is no arrow of time in conventional quantum field theory,
nor in critical string theory. Indeed, it is even possible to
formulate critical strings without introducing a time variable
at all. However, the time we experience does have an arrow, both
microscopically as codified in the second law of thermodynamics,
and macroscopically as evidenced in the cosmological Hubble
expansion. One of the main thrusts of our work has been to
understand the arrow of time in the framework of non-critical
string theory, and to relate its microscopic and macroscopic
manifestations. As we shall see, an essential feature of this
understanding is an apparent modification of quantum mechanics and
quantum field theory, entailing the abandonment of the $S$-matrix
description of scattering.
\pr
Some have long suspected that such a modification might be
necessary, in view of the fact that black holes apparently
behave thermodynamically in the context of local quantum
field theory\cite{hawk,bek,hawk2}. The appearance of an event horizon is
acompanied by non-zero entropy proportional to its area,
and a related non-zero temperature, properties that require
a mixed-state quantum treatment. This entails use of the
density-matrix formalism, and Hawking has suggested \cite{hawk}
that when space-time foam is taken into account
scattering must be
formulated as an asymptotic linear transformation from
incoming states $\rho _{in,B}^A $to outgoing states
$\rho _{out, D}^C$ :
\be
\rho _{out,D}^C = \nd{S}_{DA}^{CB} \rho _{in,B}^A
\label{one}
\ee
where the superscattering matrix $\nd{S}_{DA}^{CB}$ does not
in general factorize as a product of $S$ and $S^\dagger$ matrix
elements \cite{hawk}
\be
\nd{S}_{DA}^{CB} \ne S_A^C(S^\dagger)_D^B
\label{onetwo}
\ee
as in conventional quantum field theory. Correspondingly,
the
time-evolution of a quantum system cannot be governed
simply by the Liouville equation, which integrates to
yield just the conventional $S$ matrix, but there should
be an extra term in the quantum Liouville equation due to
space-time foam, which
we may write in the form \cite{ehns}
\be
\partial _t \rho = i[\rho, H] + \nd{\delta H}\rho
\label{two}
\ee
Such a modification of the quantum Liouville equation
is characteristic of open quantum-mechanical systems,
in which the observed (sub)system is in contact with
an unobserved reservoir. It introduces a microscopic
arrow of time, with in general
dissipation, entropy increase and apparent wave-function
collapse in the observed (sub)system.
In the quantum-gravitational
context discussed here, the unobserved states are
associated with non-trivial microscopic event horizons,
which are unobservable even in principle. We discuss below
how non-trivial contributions to the $\nd{S}$ matrix and
to $\nd{\delta H}$ may arise in string theory, using the
general formalism of string backgrounds as $\sigma$-model
field theories on the two-dimensional world-sheet. We treat the
two-dimensional (spherically-symmetric four-dimensional)
string black hole background as an illustrative example in
which specific calculations can be performed. We have
derived an explicit general form for $\nd{\delta H}$ in
string theory, expressed in world-sheet $\sigma$-model
notation and exemplified by the string black hole model.
We have also shown explicitly how this modification of
the quantum Liouville equation leads to decoherence and
apparent collapse of the wave function, as suggested
previously on the basis of more intuitive approaches to
quantum gravity.
\pr
Before discussing this, however, we discuss in section 2 some
relevant features of non-equilibrium quantum statistical
mechanics, which is the appropriate framework for the
modified quantum Liouville time-evolution equation (\ref{two}).
Specifically, we recall the general formalism of Misra and
Prigogine \cite{misra,misra2}, as well as the so-called Lie-admissible
approach of ref. \cite{santilli}, which are
compatible under certain conditions\cite{Ktorides}. In
section 3 we recall \cite{emnqm} the
general world-sheet ${\sigma}$-model derivation of the
modified quantum Liouville equation (\ref{two}), and
show that it obeys the Lie-admissibility condition of
ref. \cite{Ktorides}. This follows from the existence of
the Zamolodchikov metric in world-sheet
${\sigma}$-model coupling
space. The arrow of time is associated with
renormalization group flow in this space.
Energy is conserved in the mean \cite{emncpt},
as a consequence of renormalizability,
which replaces the time translation
invariance of conventional target-space
field theory. Probability is also conserved,
whilst entropy increases monotonically \cite{emnqm}.
Section 4 contains a more detailed discussion
of time, which is interpreted
in our approach as a renormalization group
scale identified with a Liouville field \cite{emnqm,emndollar,
emnharc}.
Section 5
reviews the string black hole model and its
interpretation in terms of monopoles \cite{emndua}, as well
as instantons \cite{yung} in this model and their
interpretation. Section 6
reviews our previous calculations of specific
contributions to the $\nd{S}$ matrix and $\nd{\delta H}$
due to monopoles and instantons on the world sheet
associated with string black holes \cite{emndua}.
In section 7 we discuss in more detail the relation
between our string black hole calculations and the
approach of ref. \cite{misra}, underlining in
particular the role played by $W$ symmetries.
Section 8 reviews the violation of $CPT$ in our
formalism, relating it to the analysis
of ref. \cite{wald} and discussing its possible
manifestation in the neutral kaon system.
In section 9 we mention applications of our approach
to cosmology, with particular mention of the initial
singularity, inflation, the time-dependences of the
fundamental parameters, and the cosmological
constant \cite{emnharc}. Finally, in section 10 we discuss the outlook
for our approach.
\pr
\section{Non-Equilibrium Quantum Statistical Mechanics
for Pedestrians}
\pr
In this section we review at an elementary level some
relevant features of non-equilibrium quantum statistical
mechanics, which we link in the next section with our
string-modified quantum Liouville equation. The
authors of \cite{misra} sought to include time
irreversibility and the second law of thermodynamics
into a description of microphysics based on the density
matrix. It is clear that this aim entails modifications
of conventional classical and quantum mechanics, that
are to be regarded as incomplete in this context.
As a first step in this programme, the authors of
\cite{misra} introduced a dynamical
transformation $\Lambda$, which is not unitary in
general,  that relates the full density
matrix $\rho$ to that for the physically-relevant
system, denoted by ${\tilde \rho}$:
\be
 {\tilde \rho}  = \Lambda\rho
\label{three}
\ee
In our case, we shall consider as ${\tilde \rho}$ the
density matrix for the low-mass light string states
that can be measured locally, as distinct from
extended solitonic string states whose properties
can only be characterized by global measurements.
We shall refer to ${\tilde \rho}$ as the locally-measurable
density matrix. Misra and Prigogine \cite{misra}
ponted out that the transformation $\Lambda$ should
satisfy certain consistency conditions: it should
preserve the positivity of $\rho$, it should obey
the equations
\be
\Lambda^. I = I
\label{four}
\ee
and
\be
\int_\Gamma d\mu Tr \Lambda \rho  = \int _\Gamma d\mu Tr \rho
\label{five}
\ee
where $\Gamma$ denotes the phase-space manifold,
and the time-evolution
operator $ U_t = e^{-iLt} $, where $L$ is the
Liouville operator, should have the intertwining
property
\be
\Lambda U_t = {{W_t}^*}\Lambda
\label{six}
\ee
for ${t \ge 0}$,
where ${{W_t}^*}$ is an adjoint strongly-irreversible Markov
semigroup operator.
\pr
Two different possibilities for $\Lambda$ should be
distinguished. One is that in which $\Lambda$ has
an inverse,
and the similarity relation
\be
{W_t}^* = \Lambda U_t \Lambda^{-1}
\label{seven}
\ee
applies for ${t \ge 0}$. This mathematical
invertibility does not mean, however,
that physical information is retained,
as we shall see later.
In the other case, $\Lambda$
has no inverse, and can be regarded as a projection
operator onto the physically-relevant states.
\pr
In the former case, which is the one that concerns us,
as we show in the next section,
the locally-measurable density
matrix ${\tilde \rho}$ obeys the time-evolution
equation
\be
i\partial _t {\tilde \rho} = \Phi (L) {\tilde \rho}
\label{eight}
\ee
where
\be
\Phi (L) = \Lambda^{-1} L \Lambda
\label{nine}
\ee
with $L$ the Liouvillian.
The existence of a function in phase space which varies
monotonically with time,
called a Lyapounov function, is guaranteed
if $\Phi(L)$ obeys the
condition \cite{misra}
\be
i\Phi (L) - i[\Phi (L)]^\dagger  \ge 0
\label{ten}
\ee
which holds if $\Lambda$
has the {\it star hermiticity} property
\be
\Lambda ^{-1} (L) = \Lambda ^{\dagger} (-L) \equiv [\Lambda (L)]^*
\label{tenb}
\ee
However, this property may not hold in general.
\pr
Time is not necessarily irreversible in this framework.
The mere existence of a Lyapounov function is
not sufficient to guarantee physical time-reversal symmetry
breaking.
Generalizing equation (\ref{seven}), one can define distinct
Markov semigroup operators
\be
{W_t}^{\pm} = {\Lambda}_{\pm} U_t \Lambda _{\pm}^{-1}
\label{eleven}
\ee
for ${t \ge 0}$ and ${t \le 0}$, respectively.
Thanks to
the time-reversal invariance
of the initial unitary evolution operator $U_t$,
time is reversible if $\Lambda_+$ = $\Lambda_-$, but
this is not the case in general. In most cases there are
physical reasons for distinguishing
$\Lambda_+ $ and $ \Lambda _{-}$, and thereby determining
the arrow of time-reversal
symmetry breaking,
as is for instance the case in which
the initial conditions in one of the Markov processes
are such that their preparation requires infinite entropy.
It has been shown \cite{misra}
that a dynamical system of the type described above
admits an internal time variable, suitable for the
discussion of aging, if $\Lambda$ obeys the conditions
(\ref{four}), (\ref{five}) and (\ref{six}),
and is sufficiently large in the sense
that for any given $\rho$ and any ${\epsilon \ge 0}$
there is another density matrix $\rho^{\prime}$ with the
property that
\be
|| \rho - U_t \Lambda \rho ' || < \epsilon
\label{twelve}
\ee
i.e., the subspace generated by $\Lambda$ evolves
backward
in
time to generate arbitrarily good approximations to
all possible states. In this case, it has been shown
\cite{misra} that one can introduce a time-evolution
operator $T$ conjugate to the Liouville operator $L$:
\be
[L,T]=i\hbar
\label{thirteen}
\ee
The converse of this theorem is also true.
\pr
In such a dynamical system, a point in phase space,
when evolved backwards in time, may become an
arbitrarily complicated and non-local region of
phase space, via a sort of inverse butterfly effect.
It is not possible to reconstruct the past history of the
system unless one measures all components of
its final-state density matrix with arbitrarily
high precision, which is impracticable. Thus, information is
effectively lost during the time-evolution, and hence
entropy increases. The system approaches an
equilibrium state in which there is no memory
of the initial state.
\pr
One example of such a system is provided by geodesics in
an expanding Universe, described by a
Robertson-Walker-Friedmann metric in ${3+1}$ dimensions:
\be
ds^2 = dt^2 - R(t)^2 \sum_{i=1}^{3}{(dx^i)}^2
\label{fourteen}
\ee
Test particles move along four-dimensional geodesics that
project onto geodesics in the three-dimensional
hypersurfaces of simultaneity. Geodesic flow in four dimensions
defines a corresponding geodesic flow in three dimensions,
and the theorem of \cite{misra2} tells us that an internal
time $T$, or age, can be defined for the system, which is not
the cosmological time, but is a monotonic function of it.
It has the interesting property that
\be
U_\lambda ^* T U_\lambda = T + \lambda (t) I
\label{fifteen}
\ee
where
\be
\lambda (t) = \lambda (t_0) + A \int _{t_0}^{t} ds \frac{1}{R(s)}
\qquad for ~ massless ~ particles
\label{sixteen}
\ee
is the
affine parameter of the projected geodesic flow, and $t$ is
the cosmic time. This tells us
that the rate of change of ${\lambda}$
becomes very rapid at early times, and very slow at
late times. Thus the rate of approach to equilibrium is
huge close to the initial singularity, and very slow in
an old Universe. This is an example of apparent
indeterminacy in the $a$ $priori$ deterministic theory
of General Relativity. As we shall see later, a very
similar situation arises in our Liouville approach to
non-critical string theory.
\pr
Thus far, we have not considered in detail the form
of the dynamical equations of motion describing the
time-evolution of such a system. An appropriate
framework is provided by the so-called Lie-admissible
formulation of dissipative statistical mechanics, as
reviewed in \cite{santilli}. The starting point that
describes the dissipative motion of a single
particle is the open version of the Lagrange equation:
\be
\frac{d}{dt} \frac{\partial L}{\partial {\dot q}^i}-
\frac{\partial L}{\partial q^i}={\cal F}_i (t,q^i, {\dot q}^i)
\label{open L}
\ee
The corresponding first-order Hamilton equations
take the form
\bea
{\dot q}^i &=& \frac{\partial H}{\partial p_i}\nn \\
{\dot p}_i &=& -\frac{\partial H}{\partial q^i} + F_i \qquad :
F_i (t,q^i, p_i) = {\cal F}_i(t, q^i, {\dot q}^i)
\label{open H}
\eea
The extension of this formulation to the statistical evolution
of the phase-space density function $\rho(q,p,t)$ entails a
generalized Liouville equation
\be
\frac{\partial \rho}{\partial t} + \{ \rho, H \}
+ F_i \frac{\partial \rho}{\partial p_i} +
\rho \frac{\partial F_i}{\partial p_i} =0
\label{genliouv}
\ee
where $\{ , \}$ denotes a conventional Poisson bracket.
As we shall see in section 3, in the string case of interest
to us the last term in this equation is absent, whilst the
previous term is very much present.
\pr
The treatment of such a system must answer the
question how the physical energy operator
$E$ can
be identified with the generator $H$ of time
translations, since ${\dot E} \ne 0$ at the operator level,
due to
dissipation or, more general, interactions with the
``environment'', whereas
\be
{\dot H} =\{ H,H \} =0
\label{hdot}
\ee
where $\{,\}$ denotes a conventional Poisson
bracket, which becomes a commutator $[,]$
in the quantum case.
The answer presented in ref. \cite{santilli}
is to modify the Lie-algebraic structure,
replacing $\{,\}$ by an object $\{\{,\}\}$
with the property that
\be
{\dot H}=\{\{H,H\}\} \ne 0
\label{doublebracket}
\ee
and analogously $[,]$ becomes $(,)$ in the quantum case.
\pr
A generalized product $(,)$ is said to form a
Lie-admissible algebra if it is linear, obeys
the generalized Jacobi identity
\bea
((A,B), C) &+& ~ cyclic ~ permutations +  \nn \\
(C, (B,A)) &+& ~ cyclic ~ permutations -  \nn \\
(A, (B,C)) &-& ~cyclic ~ permutations - \nn \\
((C,B), A) &-& ~cyclic ~ permutations = 0
\label{genjacobi}
\eea
and the condition
\be
(A,B) - (B,A) = 2 [A, B]
\label{lieadmi}
\ee
where $[,]$ denotes the conventional Lie product.
As an example, consider a dynamical system with
$\frac{\partial H}{\partial p} \ne 0$, in which case
\be
{\dot A}=\frac{\partial A}{\partial \xi^i} \omega _{ij}
\frac{\partial H}{\partial \xi^j} +
 + \frac{\partial A}{\partial \xi_i} F_i
\label{dotA}
\ee
where $\xi^i $= $(q^i,p_j)$, $\omega_{ij}=-\omega_{ji}$
is a convenient notation for the Poisson bracket commutator, and
\be
F^i = T^{ij} \frac{\partial H}{\partial \xi^j}
\label{ftij}
\ee
with
\be
T^{ij}   =\left(\begin{array}{c} 0_{n\times n} \qquad
0_{n \times n}   \\ 0_{n \times n}  \qquad s_{n \times n}
\end{array}\right) \qquad s_{ij}=\delta_{ij} (F_i/(\frac{\partial H}
{\partial p_i} ))
\label{tij}
\ee
and $0_{n \times n}$ is an $n \times n$ zero matrix.
In this case, the time-evolution equation (\ref{dotA})
can be written in the form
\be
{\dot A} = \frac{\partial A}{\partial q^i}\frac{\partial H}
{\partial p_i} - \frac{\partial A}{\partial p_i}\frac{\partial H}
{\partial q^i} + \frac{\partial A}{\partial p_i} s_{ij}
\frac{\partial H}{\partial p_j} = (A, H)
\label{dotS}
\ee
It is easy to check that the product $(,)$ defined
in this equation is of the Lie-admissible form
defined earlier, and reduces to the conventional
Lie product $[,]$ if $s_{ij}$ = $0$.
\pr
We are now in a position to compare this Lie-admissible
formulation of dissipative statistical mechanics with
the previous formulation of \cite{misra}, based on a
non-unitary time-evolution operator $\Lambda$. We
recall the form (\ref{eight}) of the time-evolution
equation for the locally-measurable density matrix
${\tilde \rho}$ in that formulation, and the star
hermiticity condition (\ref{ten},\ref{tenb}). For the type
of dynamical system described by (\ref{open L},\ref{open H},
and \ref{dotS}), we have
\be
\Phi =i\sum _{i=1}^{n} (\frac{\partial H}{\partial q^i}
\frac{\partial }{\partial p_i} - \frac{\partial H}{\partial p_i}
\frac{\partial}{\partial q^i}) + \sum_{i,j=1}^{n}
\frac{\partial H}{\partial p_i} s_{ij} \frac{\partial}{\partial p_j}
\label{PhiCK}
\ee
which is consistent with star-hermiticity if and only if
\be
 {\cal F} = {\cal F}^\dagger \qquad; \qquad {\cal F} \equiv
\sum_{i,j=1}^{n}\frac{\partial H}{\partial p_i} s_{ij}
\frac{\partial}{\partial p_j}
\label{CK}
\ee
A necessary and sufficient condition that this
be satisfied is that the matrix $s_{ij}$ be real
and symmetric \cite{Ktorides}. Dissipative statistical systems
with this property provide examples of the
$\Lambda$-transformation theory of \cite{misra}.
As we shall see in subsequent sections, string theory
is one such example.
\pr
\section{Review of String Density Matrix Mechanics}
\pr
In order to accommodate the mixed states inevitable in
a quantum theory of gravity, one must use a
density matrix formalism \cite{hawk,ehns}.
The asymptotic $\nd{S}$ matrix may not have the familiar analyticity
properties, and may only exist in a distribution-theoretic sense,
as we shall see later. The evolution of the density matrix
over finite times is given by a
modification (\ref{two})
of the quantum Liouville equation \cite{ehns}.
The commutator term in (\ref{two}) would, when
integrated alone over time, give the conventional
$S$ matrix. The extra term $\nd{\delta H}$ leads, when
integrated over time, to the non-factorization property
$\nd{S} \ne S S^\dagger $. As seen in section 2,
the presence
of such an extra term is characteristic of a
dissipative quantum-mechanical system such as
an open
system interacting with an unobserved
environmental reservoir \cite{vernon,cald,davydov}. In
string density matrix mechanics
this term would reflect the mixing of observable particles
to unobservable quantum gravitational states, whose
information is lost across microscopic event horizons.
Such states are delocalized, and can be thought of as remnants
of a symmetric (topological) phase of gravity \cite{emntop},
whose breaking might result in the emergence of
the ordinary space-time.
\pr
How should one set about modelling space-time foam and
exploring
these possibilities in string theory?
In order for the background in which it
propagates have a classical space-time interpretation,
the string theory must be critical, i.e. must be
characterized by a conformal field theory on the world sheet
with
central charge $c=26$ ($15$) for a bosonic (supersymmetric) string.
It is well known \cite{stringbook} how to reproduce $S$-matrix
elements via the operator product expansion for such a
critical-string conformal field theory. One must look
beyond this framework if one is to have any chance of
locating a non-trivial contribution to the $\nd{S}$ matrix.
This means looking at quantum fluctuations in the space-time
background, and in particular transitions between them.
The appropriate space of theories to discuss
these is that of two-dimensional field theories on the
world-sheet, but without the restriction that these
be conformal. This is the space of generalized $\sigma$-models
on the world sheet that we have used to
derive \cite{emnshort,emndollar}
a form for ${\nd{\delta H}}$ in the framework of non-critical
string theory\cite{aben3}.
\pr
Such models are characterized by deformations
of conformal points, described
by vertex operators  $V_i$
associated
with background fields $g_i$ in the target space
in which the string propagates.
Mathematical consistency requires
the
restoration of
criticality by turning a
deformation
into an exactly marginal one by Liouville dressing, thereby
leading, in our interpretation,
to a time-dependent background.
To be more explicit,
consider a two-dimensional
critical conformal field theory model
described by an action $S_0(r)$ on the world sheet,
where the
$\{ r \}$ are matter fields
spanning a $D$-dimensional target manifold  of Euclidean
signature, that we
term
``space''.  Consider, now, a
deformation
\be
    S=S_0 (r) + g \int d^2 z V_g (r)
\label{nonmarg}
\ee
Here $V_g$
 is a (1,1) operator,  i.e. its anomalous dimension vanishes, but
it is not {\it exactly marginal}
in the sense that the operator-product
expansion coefficients $C_{ggg}$
of $V_g$ with itself are non-zero in any renormalization
scheme,
and hence
`universal' in the Wilsonian sense.
The scaling dimension $\alpha _g$
of $V_g$
in the deformed theory (\ref{nonmarg}) is, to $O(g)  $ \cite{ginsp},
\be
           \alpha _g =-g C_{ggg} + \dots
\label{scaling}
\ee
Liouville theory \cite{DDK}
requires  that scale invariance
of the theory (\ref{nonmarg}) be preserved. The non-zero
scaling dimension
(\ref{scaling}) would jeopardize
this, but
scale invariance
is
restored if
one dresses $V_g$ gravitationally on the world-sheet as
\be
\int d^2z g V_g (r)
\rightarrow
 \int d^2z g e^{\alpha _g \phi}V_g (r)
 =\int d^2z g V_g (r)
 -\int d^2z g^2 C_{ggg}V_g (r)\phi  + \dots
\label{dressed}
\ee
where $\phi$ is the Liouville field. The latter acquires dynamics
through integration over world-sheet covariant metrics
$\gamma _{\alpha\beta}$
after
conformal gauge fixing $\gamma _{\alpha\beta}=e^\phi {\hat
\gamma_{\alpha\beta}}$
in the way discussed in \cite{DDK,mm}.
Scale invariance is guaranteed through the definition
of renormalized couplings $g_R$ , given in terms of $g$ through
the relation
\be
    g_R\equiv g - C_{ggg}\phi g^2  + \dots
\label{renorm}
\ee
This
equation leads to the correct $\beta$-functions for $g_R$
\be
      \beta _g =-C_{ggg}g_R^2  + \dots
\label{beta}
\ee
implying a renormalization-group scale $\phi$-dependence of
$g_R$. The reader
might have noticed that above
we viewed
the Liouville field as a local scale on the world-sheet
\cite{emnqm,emndollar}. Local world-sheet scales
have
been
considered in the past
\cite{shore,osborn}, but the crucial difference in our Liouville
approach is that
this scale is made
dynamical  by being
integrated out
in the
path-integral. In this way, in Liouville strings the local dynamical
scale acquires
the interpretation of an additional target
coordinate. If the central charge of the matter theory is $c_m > 25$,
the signature of the kinetic term of the Liouville  coordinate
is opposite to that of the matter fields $r$, and thus the Liouville
field is interpreted as Minkowski target time \cite{aben3,aben},
as we discuss in more detail in the next section and in
ref. \cite{emndollar,emnharc}.
\pr
The target-space density
matrix in such a framework
is viewed  as a function of coordinates
$g^i$
that parametrize the couplings of the generalized
$\sigma$-models on the string world-sheet, and their
conjugate momenta $p_i$ : $\rho (g^i, p_i)$.
The effective action functional for this dynamical system
can be
identified with the Zamolodchikov $C$-function \cite{zam}
$C (g^i)$, whose gradient determines the rate of
change of the coordinates (couplings) $g^i$ :
\be
{\dot g}^i = \beta ^i(g) \qquad : \qquad
\frac{\delta    C (g)}{\delta   g^i} =G_{ij}\beta ^j
\label{threet}
\ee
where
\be
C    [g] =\int dt (p_i {\dot g}^i - E )
\label{Fef}
\ee
where $E$ is the Hamiltonian
and
$G_{ij}$ is the metric in $g$-space,
\be
G_{ij}[g]=2 |z|^4 <V_i(z)V_j(0)>
\label{zametric}
\ee
with $V_i$ the associated vertex operators corresponding
to the background $g^i$ and $< \dots >$ denotes a
$\sigma$-model vacuum expectation value.
Here and subsequently dots denote derivatives
with respect to the renormalization scale.
As
described above and in ref. \cite{emnharc}
and the next section, we identify the
renormalization scale with a Liouville field,
which has negative metric in target space
because fluctuations make the string supercritical
\cite{aben3}, and we identify it with the target time
variable. This is the main reason
for considering in (\ref{Fef}) an `effective action'
and not simply a Lagrangian in target space. This
is crucial in our formalism, as a result
of the integration over the dynamical Liouville scale
in a $\sigma$-model
path integral. In this formalism, the simple
gradient flow of the off-shell corollary of the
$C$-theorem \cite{zam,osborn,santos}
is extended to a non-trivial functional
derivative
\be
\frac{\delta }{\delta g^i} C[g]
=-\frac{d}{dt} \frac{\partial
 \cal L(g,{\dot g}, t)}{\partial {\dot g^i}} + \frac{\partial {\cal L}}
{\partial g^i}
\ne 0
\label{gradient}
\ee
which includes a generalized non-potential
force term in  the evolution equation for $g^i$ \cite{emnqm}
derived from (\ref{Fef}). The renormalizability of the world-sheet
$\sigma$-model implies
\be
 \frac{d}{d t} \rho [g^i,p_i, t] = 0 =
 \frac{\partial }{\partial t} \rho
  + {\dot g}^i \frac{\partial }{\partial g^i} \rho
+ {\dot p}_i \frac{\partial }{\partial p_i} \rho
\label{exisosh}
\ee
 From (\ref{gradient}) one can then derive straightforwardly
a modified Liouville equation (\ref{two})
with the explicit form \cite{emnqm}
\be
\nd{\delta H}\rho =G_{ij}\beta^j \frac{\partial \rho}
{\partial p_i}=-iG_{ij}[\rho, g^i]\beta^j
\label{fourt}
\ee
where the second form holds in the quantum formulation.
\pr
We note here the fundamental point that this
modification of the quantum Liouville equation
obeys  the Lie-admissibility condition
of ref. \cite{Ktorides}, because the
Zamolodchikov tensor (\ref{zametric})
is real and symmetric, and hence defines
a metric in coupling constant space.
This property is non-trivial,
since it does not hold in a
a more general renormalization scheme
 \cite{osborn}, in which
the couplings $g^i$ have an arbitrary
dependence on the world-sheet coordinates
$g^i(\sigma, \tau)$ and not a simple
local scale dependence \cite{osborn}.
In such a scheme
there exists a local (in
renormalization group space) function
whose variations with respect to the couplings
have off-shell relations with the $\beta$-functions given by
matrices that have antisymmetric parts.
More explicitly,
one can prove the relation \cite{osborn}:
\be
\partial _i {\tilde \beta}^\Phi =\chi _{ij} \beta^j
+\partial _j W_i \beta^j - \partial _i W_j \beta^j
\label{betatwidl}
\ee
where $\Phi$ is the dilaton background coupled to the
world-sheet curvature, the $g^i$ denote the rest of the backgrounds,
existing even on flat world sheets, and
$\chi _{ij}$ is symmetric. It is
related to divergences of the
two-point functions of $V_i$,
but its positivity
is not evident in this approach. Finally,
${\tilde \beta}^\Phi \equiv \beta^\Phi
+ W_i \beta^i $, where the
$W_i $ are (computable)
renormalization
counterterms related to total world-sheet
derivative terms in the expression for the
trace of the stress-tensor. Such terms are
crucial for the local scale invariance of the theory.
In dimensional regularization the
$W_i$ are non-trivial
beyond three $\sigma$-model loop order \cite{osborn}.
Such off-shell relations appear only
if the
$g^i$ are allowed to depend arbitrarily
on the world-sheet coordinates, which
is not the case in the framework adopted above
and in ref. \cite{shore}
where the $g^i$ depend only on the renormalization
scale that we identify with target time \cite{emnqm,emndollar}.
\pr
The more general framework incorporates
off-shell generalized forces that depend
on the conjugate momenta $p_i$. It
could be regarded as providing
an `atlas' relating `charts' or
`patches', in each of which
the $g^i$ space is torsion-free. In our
interpretation, the Universe observable
within our event horizon is contained
within one patch, enabling physics
everywhere within it to be described by the approach
to a unique conformal field theory
or critical string vacuum. There may
well be other patches beyond our
observable horizon, in which
a different string
vacuum is approached,
with the generalized
renormalization scheme
\cite{osborn}, including torsion
(\ref{betatwidl}) describing
transitions between the patches.
This would provide a physical
realization of the rather
abstract analysis in ref. \cite{vafa},
where it was argued that
the `classical' coupling constant space
should not be simply connected, with
each component corresponding to one of our
patches.
\pr
This `torsion-free'
modification of the quantum Liouville
equation, that applies within our patch of the
Universe,
has other several important properties.
One is that the total probability
$P=\int dp_l dg^l Tr[\rho(g^i, p_j)]$ is
conserved :
\be
{\dot P}=\int dp_l dg^l Tr[\frac{\partial }{\partial p_i}
(G_{ij}\beta^j \rho )]
\label{fivet}
\ee
which can receive contributions only from the boundary of phase space,
that must vanish for an isolated system.
Secondly, energy is conserved on the average \cite{emncpt}.
This can be seen by computing
\bea
\partial _t <<E^n>> =n<<(\partial _t E)E^{n-1}>>-i<<\beta^i G_{ij}
[g^j, E^n]>>
= \nn \\
n<<(\partial _tE) E^{n-1}>>-i<<\beta^iG_{ij}
E[g^j,E^{n-1}]]>>+<<\beta^i G_{ij} \beta^j E^{n-1}>>
\label{fluctuations}
\eea
where $E$ is the Hamiltonian operator, and
$<< \dots >> \equiv Tr[\rho (\dots)]$.
In arriving at this result we took into account the
quantization rules in
coupling constant space discussed in ref. \cite{emnqm},
\be
[g^i, g^j]=0 \qquad ; \qquad [g^i, p^j]=-i\delta^{ij}
\label{ccs}
\ee
as well as the fact that in string $\sigma$-models
the `quantum operators'
$\beta^i G_{ij}$ are
functionals of the coordinates only $g^i$
and not of the generalized momenta $p^i$. For future use we note that
the total time derivative of an operator ${\hat Q}$ is given as
usual by
\be
\frac{d}{dt} {\hat Q} =-i[ {\hat Q}, E]
\label{timederivative}
\ee
We recall that total time derivatives
incorporate both explicit and implicit (via running
couplings) renormalization-scale dependence, whilst
partial time derivatives incorporate only the explicit dependence.

\pr
For the energy conservation law
we should take $n=1$ in (\ref{fluctuations}), in which case we find
\be
\frac{\partial}{\partial t}
<<E>>=
\frac{\partial}{\partial t} Tr(E\rho) = <<\partial _t (E -
\beta^i G_{ij} \beta^j)>>
\label{energcons}
\ee
Using
the $C$-theorem
results  (\ref{threet},\ref{Fef})
\cite{zam,emncpt}
and the formalism developed in ref. \cite{emnqm}
it is straightforward to arrive at
\be
\frac{\partial}{\partial t}
<<E>>=
\frac{\partial}
{\partial t} (p_i\beta^i) = 0
\label{sixt}
\ee
due to the renormalizability of the stringy $\sigma$-model.
The latter
implies that any dependence on the renormalization
group scale in the $\beta^i$ functions
is implicit through the renormalized couplings.
Renormalizability replaces the time-translation
invariance of conventional target-space field theory.
\pr
This conservation
result does not generalize to
quantum fluctuations in the energy $\Delta E =<<E^2>>-(<<E>>)^2$.
To get the
the energy fluctuations we set $n=2$,
\be
\partial _t <<E^2>>=-i<<[\beta ^j, E]
\beta ^i G_{ij}>>
=<<\frac{d \beta ^j}{d t}\beta^i G_{ij}>>
\label{de}
\ee
which is non-zero for a non-critical string.
This result implies that despite energy conservation
the uncertainties in the energy $\Delta E$ depend on the
(Liouville) time $t$. This point is relevant to the
discussion of uncertainty in section 9 \cite{emnharc}.
\pr
The entropy
$S=-Tr(\rho ln \rho)$ is also
not conserved:
\be
{\dot S}=(\beta^i G_{ij} \beta^j)S
\label{sevent}
\ee
implying a monotonic increase for unitary theories
for which $G_{ij}$ is positive definite. We see from (\ref{sevent})
that {\it any} running of {\it any} coupling will lead to an
increase in entropy, and we have interpreted \cite{emnqm} this
behaviour in terms of quantum models of friction \cite{cald}. The
increase (\ref{sevent}) in the entropy corresponds to a
loss of quantum coherence, which is also known in these
models. Note that entropy increases within any `torsion-free'
cosmological `patch' in coupling space: this is not in general true
at the boundaries between patches, where `torsion' may appear.
\pr
The final comment in this brief review of density
matrix mechanics is that Ehrenfest's theorem continues
to hold. The time evolution of the
expectation value of any observable $O(g^i)$
that is a function of the coordinates alone, and not the
momenta $p_i$, is given by
\bea
\frac{\partial}{\partial t}
<<O(g^i)>>=
\frac{\partial}{\partial t} Tr(O(g^i)\rho) =
Tr(O(g^i){\dot \rho}) \nn \\
=iTr[O(g^i)[\rho, H]]+iTr[g^i,O(g^i)\beta^iG_{ij}\rho]=
iTr\{O(g^i)[\rho, H]\}
\label{thirteent}
\eea
as usual.
\pr
It may be helpful to bear in mind a simple
two-state system, whose $2 \times 2$ density matrix
can be decomposed with respect to the hermitian
Pauli $\sigma$-matrix basis \cite{ehns}
$({\bf 1}, \sigma _x, \sigma_y, \sigma _z)$,
$\rho =\rho _0 {\bf 1} + {\bf \rho}^. {\bf \sigma}$.
In conventional quantum mechanics with a Hamiltonian
$H=\Delta E \sigma _z $, a pure initial state
$\rho _{in} =\frac{1}{2} (|1> + |2>)(<1| + <2|) $ evolves
unitarily :
\be
\rho (t) =\frac{1}{2}\left(\begin{array}{c} 1 \qquad
e^{-i\Delta E t} \\ e^{i\Delta E t} \qquad 1 \end{array}\right)
\label{eightt}
\ee
A generic ``open system'' modification $\nd{\delta H}$
can be expressed as a $4 \times 4$ matrix w.r.t. the coordinates
$(0, \sigma_x, \sigma_y, \sigma_z)$. The
probability and energy conservation
derived above (\ref{fivet},\ref{sixt}) tell us that
\be
\nd{\delta H}_{0\beta}=0=\nd{\delta H}_{\beta 0} , \qquad
\nd{\delta H}_{3\beta} =0 =\nd{\delta H}_{\beta 3}
\label{ninet}
\ee
respectively. We can therefore write
\be
\nd{\delta H}_{\alpha \beta} =\left(\begin{array}{c}
0 \qquad 0 \qquad 0 \qquad 0 \\
0 \qquad -\alpha \qquad -\beta \qquad 0 \\
0 \qquad -\beta \qquad -\gamma \qquad 0 \\
0 \qquad 0 \qquad 0 \qquad 0 \end{array}\right)
\label{tent}
\ee
where
positivity imposes  $\alpha, \gamma > 0, \alpha \gamma > \beta ^2 $.
It is easy to see that with the addition of such a term
\be
\rho (t) = \frac{1}{2} \left(\begin{array}{c}
1 \qquad e^{-(\alpha + \gamma)t/2}e^{-i\Delta E t} \\
e^{-(\alpha + \gamma)t/2}e^{i\Delta E t} \qquad 1 \end{array}\right)
\label{elevent}
\ee
which becomes asymptotically a completely mixed state
\be
\rho (\infty) =\frac{1}{2} \left(\begin{array}{c}
1 \qquad 0 \\
0 \qquad 1 \end{array}\right)
\label{twelvet}
\ee
Such evolution towards a completely mixed state
is the generic consequence of the monotonic increase
(\ref{sevent}) in the entropy. The completely-mixed
form (\ref{twelvet}) of the density matrix corresponds
to the advertized loss of coherence.
\pr
As discussed in more detail in section 8,
the above formalism can be adapted, with
conceptually minor modifications, to accommodate
decays to the neutral $K^0-{\overline K}^0$ system
\cite{ehns,emncpt}, which is one
of the best microscopic laboratories for testing
quantum mechanics and its possible modification (\ref{two}).
The coresponding parameters $\alpha, \beta , \gamma$
(\ref{tent}) violate $CPT$. String theory suggests
that $CPT$ violation should be considered  a generic
feature of our density matrix mechanics \cite{emncpt}. The
normal
field-theoretical proof of the $CPT$ theorem
is based on locality, Lorentz invariance and
unitarity. Clearly string theory is not local in
space-time, and Lorentz invariance
may be considered {\it
a derived property of critical string
theory that does not hold in our treatment of time as a
renormalization scale in non-critical string theory}.
Indeed, we have related $CPT$ violation
to charge non-conservation on the world-sheet associated
with topological fluctuations such as monopoles whose appearance
drives the string supercritical \cite{emncpt,emndollar}.
\pr
\section{Time and the Two-Dimensional String Black Hole Model}
\pr
As an illustration of our approach to non-critical string theory,
we now discuss the two-dimensional black hole model of ref. \cite{witt}.
We
regard it as a toy laboratory that gives us insight into the nature
of time in string theory and contributes to the physical effects
mentioned in the previous section.
\pr

The action of the model is
\be
   S_0=\frac{k}{2\pi} \int d^2z [\partial r {\overline \partial } r
- tanh^2 r \partial t {\overline \partial } t] + \frac{1}{8\pi}
\int d^2 z R^{(2)} \Phi (r)
\label{action}
\ee
where $r$ is a space-like coordinate and $t$ is time-like,
$R^{(2)}$ is the scalar curvature, and $\Phi$ is the dilaton field. The
customary interpretation of (\ref{action}) is as a string model with
$c$ = 1 matter, represented by the $t$ field, interacting with a
Liouville mode, represented by the $r$ field, which has $c  < 1$ and
is correspondingly space-like \cite{aben3,aben}.
As an illustration of the
approach outlined in the previous section, however, we re-interpret
(\ref{action}) as a fixed point of the renormalization group
flow in the local scale variable $t$. In our interpretation, the
``matter'' sector is defined by the spatial coordinate $r$, and has
central charge $c_m$ = 25 when $k  = 9/4$
\cite{witt}. Thus the model
(\ref{action}) describes a critical string in a dilaton/graviton
background. The fact that this is static, i.e. independent of $t$,
reflects the fact that one is at a fixed point of the renormalization
group flow \cite{emndollar,emnharc,emnnew}.
\pr
We now outline how one can use the machinery
of the renormalization group in curved space,
with $t$ introduced as
a local renormalization
scale on the world sheet, to derive the model (\ref{action}). A detailed
technical description is given in \cite{emnnew,emndollar}.
There are two contributions
to the kinetic term for $t$ in this approach, one associated with
the Jacobian of the path integration over the world-sheet metrics, and
the other with fluctuations in the background metric.
\pr
To exhibit the former, we first choose the conformal gauge
$\gamma _{\alpha \beta}=e^{\rho}
{\hat \gamma}_{\alpha\beta}$ \cite{DDK,mm},
where $\rho$ represents the Liouville mode. We will later identify $\rho$
with an appropriate function of $\phi$, thereby making the local scale
$\phi$ a dynamical $\sigma$-model field. Ref.\cite{mm}
contains an
explicit computation of the Jacobian using heat-kernel regularization,
which yields
\be
-\frac{1}{48\pi}[\frac{1}{2}
\partial_\alpha \rho \partial^\alpha \rho +
R^{(2)}\rho + \frac{\mu}{\epsilon} e^\rho +
S'_G ]
\label{liouvillekin}
\ee
where the counterterms $S'_G$ are needed to remove the non-logarithmic
divergences associated with the induced world-sheet
cosmological constant term $\frac{\mu}{\epsilon}e^\rho$,
and depend on the background fields. This procedure
reproduces the critical string results of ref. \cite{witt}
when one identifies
the Liouville field $\rho$ with 2$\alpha'  \phi$. Equation
(\ref{liouvillekin})
contains a negative (time-like) contribution to the kinetic term
for the Liouville (time) field, but this is not the only such
contribution, as we now show.
\pr
We recall that the renormalization of composite operators in
$\sigma$-models formulated on curved world sheets is achieved by
allowing an arbitrary dependence of the couplings $g^i$ on the
world-sheet variables $z,{\bar z}$ \cite{shore,osborn}.
This induces counterterms of ``tachyonic'' form,
which take the following form in dimensional regularization with
$d  = 2 -  \epsilon$ \cite{osborn}:
\be
\int d^2z \Lambda_0
\label{tachloc}
\ee
where
\be
\Lambda _0=\mu^{-\epsilon}(Z(g)\Lambda + Y(g))
\label{ctr}
\ee
Here
$Z(g)$ is a common
wave function renormalization that maps target scalars into scalars,
$\Lambda$ is a residual
renormalization factor,
and the remaining counterterms $Y(g)$ can be expanded as power series
in 1/$\epsilon$, with the the one-loop result giving a simple pole.
Simple power-counting yields the following form for $Y(g)$:
\be
Y(g)=\partial _\alpha g^i {\cal G}_{ij} \partial ^\alpha g^j
\label{zamolmetric}
\ee
where ${\cal G}_{ij}$
is the analogue of the Zamolodchikov metric
\cite{zam} in this formalism, which is
positive for unitary
theories. It is related to the divergent part of the
two-point function $<V_i V_j>$ \cite{osborn} that
cannot be absorbed
in the conventional renormalization of the operators $V_i$. We need
to consider a $\sigma$-model propagating in a graviton background
$G_{MN}$, in which case a standard one-loop computation \cite{osborn}
yields the
following result for the simple $\epsilon$-pole in $Y$:
\be
Y^{(1)}=\frac{\lambda}{16\pi \epsilon}\partial _\alpha G_{MN}
\partial^\alpha G^{MN}
\label{polemetric}
\ee
where $\lambda\equiv 4\pi \alpha '$ is a loop-counting parameter.
We note that the wave-function renormalization
$Z(g)$ vanishes at one-loop.
In ref. \cite{osborn}
$G_{MN}$ was allowed to depend arbitrarily on the world-sheet
variables, and all world-sheet derivatives of the couplings were
set to zero at the end of the calculation. In our Liouville mode
interpretation, we assume that such dependence occurs only through
the local scale
$\mu (z,{\overline z})$,
so that
\be
\partial _\alpha g^i ={\hat \beta^i}\partial _\alpha \phi (z,{\bar z})
\label{part}
\ee
where ${\hat \beta }^i=\epsilon g^i + \beta ^i(g)$
and $\phi=ln \mu (z,{\bar z})$.
Taking
the $\epsilon \rightarrow 0$ limit, and
separating the finite and $O(\frac{1}{\epsilon})$ terms,
we obtain for the former
\be
O(1)-terms : \qquad
Res Y^{(1)}=\alpha '^2 R \partial_\alpha \phi
\partial^\alpha \phi
\label{scalarcurv}
\ee
where $R$ is the scalar curvature in target space, and we have used
the fact that the one-loop graviton $\beta$-function
is
\be
\beta _{MN}^G=\frac{\lambda}{2\pi}R_{MN}
\label{gravbeta}
\ee
The terms without logarithmic divergences,
\be
      \frac{1}{\epsilon} \beta^i {\cal G}^{(1)}_{ij} \beta^j
\label{quadratic}
\ee
do not contribute to the renormalization group, and can be
removed explicitly by target-space metric counterterms
\be
S_G =\frac{1}{\epsilon} G_{\phi \phi} \partial _\alpha \phi
\partial^\alpha \phi
+ \delta S(\phi, r)
\label{metric}
\ee
where the coefficients $G_{\phi\phi}$ are
fixed by the requirement
of cancelling the $\frac{1}{\epsilon}$ terms (\ref{quadratic}).
The $\delta S$ denotes arbitrary finite counterterms, which
are invariant under the simultaneous
conformal
rescalings of the fiducial world-sheet metric, ${\hat \gamma}
\rightarrow
e^{\sigma}{\hat \gamma}$, and local
shifts of the scale $\phi \rightarrow
\phi -\sigma$. This last requirement arises as in
the conventional approach to Liouville gravity \cite{DDK,mm}, where
the local renormalization scale $\phi$ is identified with the
Liouville mode $\rho$, after appropriate normalization. In our
interpretation
one is forced to treat the scale $\phi$ simultaneously as
the target time coordinate.
\pr
In the case of the Minkowski black hole model of ref. \cite{witt},
the
Lorentzian curvature is
\be
R=\frac{4}{cosh^2r}=4-4tanh^2r,
\label{curba}
\ee
which we substitute into equation (\ref{scalarcurv})
to obtain the form of the
second contribution to the kinetic term for the Liouville field $\phi$.
Combining the world-sheet metric Jacobian term in (\ref{liouvillekin})
with the
background fluctuation term
(\ref{scalarcurv},\ref{curba}),
we finally obtain the
following terms in the effective action
\be
\frac{1}{4\pi \alpha '} \int d^2z [
\partial _\alpha r \partial ^\alpha r -
tanh^2r \partial_\alpha \phi
\partial^\alpha \phi + dilaton-terms ]
\label{effectact}
\ee
Thus we recover the critical string $\sigma$-model action
(\ref{action})
for the Minkowski black-hole.
Dilaton counterterms are incorporated
in a similar way, yielding the dilaton
background of \cite{witt}. In addition,
as standard in stringy
$\sigma$-models, one also obtains the necessary
counterterms that guarantee target-space diffeomorphism
invariance of the Weyl-anomaly cefficients \cite{shore}.
Details are given in ref. \cite{emnnew}.
\pr
It should be noticed that
the renormalization group yields automatically the Minkowski
signature, due to the $c_m =25$ value of the matter central charge
\cite{aben3,aben}.
However,
as we
remarked in ref. \cite{emnnew,emndollar},
one can also switch over
to the Euclidean black hole model, and still maintain the
identification of the compact time with some appropriate function
of the Liouville scale $\phi$  that takes into account the
compactness of $t$ in that case.
The formalism of
exactly-marginal
deformations that turn on matter in the model (\ref{action})
is better studied in this Euclidean version \cite{chaudh}.
In ref. \cite{chaudh} it was argued that the exactly-marginal
deformation that turned on a static tachyon background for the
black hole of ref. \cite{witt} necessarily involved the
higher-level topological string
modes, that are non-propagating
delocalized
states, which are interrelated by an infinite-dimensional
$W$ symmetry\footnote{The elevation of this symmetry
to target space-time is discussed in more detail
in section 7.}. This is a consequence
of the operator product expansion of the tachyon zero-mode operator
${\cal F} _{-\frac{1}{2},0} ^c    $ \cite{chaudh}:
\be
    {\cal F} _{-\frac{1}{2},0}^c    \circ
    {\cal F} _{-\frac{1}{2},0}^c    = {\cal F}_{-\frac{1}{2},0}^c
+ W_{-1,0}^{hw} + W_{-1,0}^{lw} + \dots
\label{ope}
\ee
where we only exhibit
the
appropriate holomorphic part for reasons of economy of
space.
The
$W$ operators and the $\dots$ denote level-one and higher string
states.
The corresponding exactly-marginal deformation,
constructed by tensoring holomophiic and  antiholomorphic parts,
is given by \cite{chaudh}
\be
L_0^1{\overline L}_0^1 \propto
{\cal F}^{c-c}_{\frac{1}{2},0,0} + i(\psi^{++}-\psi^{--}) + \dots
\label{margintax}
\ee
where the $\psi$ denote higher-string-level operators \cite{chaudh},
and
the `tachyon' operator is given by
\be
{\cal F} ^{c-c}_{\frac{1}{2},0,0}(r)
=\frac{1}{coshr}
F(\frac{1}{2},\frac{1}{2} ; 1, tanh^2r )
\label{tachyon}
\ee
with
\bea
&~&F(\frac{1}{2},\frac{1}{2},1;tanh^2r) \simeq
\frac{1}{\Gamma ^2(\frac{1}{2})}\sum_{n=0}^{\infty}
\frac{(\frac{1}{2})_n(\frac{1}{2})_n}{(n !)^2}[2\psi(n+1)-
2\psi(n+\frac{1}{2})+ \nn \\
&+&ln(1 + |w|^2)]
(\sqrt{1 + |w|^2}~)^{-n}
\label{wseven}
\eea
There is an additional marginal deformation, dictated by the
$SL(2,R)$ symmetry structure \cite{chaudh}, which consists
of topological string modes only.
At large
$k$, this  operator  rescales the black hole metric, as
can be seen
from
its contribution to the action of the deformed
Wess-Zumino
$\sigma$-model after the gauge field integration \cite{chaudh},
\bea
gL_0^2 {\overline L_0}^2 \ni \int d^2z \{\partial r {\overline
\partial} r (1-2g csch^2 r -2g sech^2 r) + \nn \\
\partial \theta {\overline \partial}\theta
(sinh^2 r + 2g - \frac{(sinh^2 r + 2g)^2}{cosh^2 r + 2g})\}
\label{chlyk}
\eea
Changing variables $cosh^2r + 2g \rightarrow cosh^2r $ in (\ref{chlyk})
one finds that to $O(g)$ the target space metric is rescaled by an
overall constant.
\pr
The topological (higher-level) string modes
cannot be detected in local scattering
experiments, due to their delocalized character.
 From a formal field-theoretic point of view, such states cannot
exist as asymptotic states to define scattering, and also cannot
be integrated out in a local path-integral.
An `experimentalist' therefore sees necessarily a
truncated matter theory, where the only deformation  is the
tachyon ${\cal F}_{-\frac{1}{2},0}^c    $, which is a (1,1) operator
in the black hole $\sigma$-model (\ref{action}), but is not
{\it exactly} marginal. This truncated theory
is non-critical, and hence
Liouville dressing in the sense of
(\ref{dressed})
is essential, thereby implying
time-dependence of the matter background.
Due to the
fact that the
appropriate
 exactly-marginal deformation associated with the tachyon in
these models
 involves all the higher-level
string states that are interrelated by $W$-symmetry, one can
conclude that in this picture the ensuing
non-equilibrium time-dependent backgrounds
are a consequence
of information carried off
by the unobserved topological string modes.
These
states are delocalized modes with definite (target-space)
energies and momenta, so a
low-energy
scattering process involving
propagating string degrees of freedom
will not have any observable
energy violations.
This is an apparent
physical explanation in this case for the
general result
(\ref{sixt}) of
energy conservation on the average in density matrix mechanics.
The r\^ole of the space-time singularity\footnote{We would like to
stress that the notion of `singularity' is clearly a low-energy
effective-theory concept. The existence of infinite-dimensional
stringy symmetries associated with higher-level string states
($W_\infty$-symmetries \cite{emn1}) `smooth out' the singularity,
and render the full string theory finite.}
 was crucial for this argument.
Indeed, in flat target-space matrix models \cite{matrix} the
tachyon zero-mode operator ${\cal F}$ is exactly marginal. As we shall
argue later on,   these flat models can be regarded as
an asymptotic ultraviolet
limit in time
of the Wess-Zumino black hole. Hence,
any time-dependence of the matter disappears in the vacuum,
leading to equilibrium.
\pr
The above `truncation' procedure can be compared
with the $\Lambda$ transformation theory of Misra and
Prigogine \cite{misra}, and its Lie-admissible
formulation \cite{santilli},
discussed in section 2.
It has been argued in ref. \cite{emnwhair}
that the `topological' modes can in principle be observed
by either Aharonov-Bohm (global) scattering experiments,
in four-dimensional string theories, or via selection rules
characterizing tachyon scattering off black-hole backgrounds
in the (effective) two-dimensional ($s$-wave four-dimensional)
string theory, in an analogous fashion to fermion
scattering on a monopole in the Callan-Rubakov
effect \cite{callan}. This, as well as the coherence-maintaining
property of the associated $W_\infty$-algebras that characterize the
two-dimensional string theory \cite{emn1},
imply that no information is lost
in the $\Lambda$ transformation and the
associated $\Lambda$ operator is therefore invertible,
as we discuss in more detail in section 7.
The Lie-admissible structure then is evident from the
work of \cite{Ktorides} and the reality of the
Zamolodchikov metric tensor $G_{ij}$ (\ref{zametric}).
The modified Liouville equation of string density matrix
mechanics (\ref{fourt}) is to be compared to
the (quantum version) of the generalized Liouville equation
(\ref{genliouv}) or the equivalent Misra-Prigogine form
(\ref{PhiCK}). The $C$-theorem of Zamolodchikov \cite{zam},
as extended in section 3 to incorporate Liouville strings,
provides us with a natural Lyapounov function in the
coupling constant (background field) phase space, as
discussed in sections 2 and 3.
\pr
Having such a Lyapounov function,
it is natural to enquire into the irreversibility
in target time of the  effective theory of locally-measurable
observables. To this end, we recall that in
Liouville theory a correlation function of $(1,1)$ matter
deformations $V_i$ is given by \cite{Li}
\be
< V_{i_1} \dots V_{i_n} >_\mu = \Gamma (-s)
< V_{i_1} \dots V_{i_n} >_{\mu=0}
\label{Liouvcorr}
\ee
where $s$ is the sum of the appropriate Liouville energies,
and $<\dots >_\mu $ denotes a $\sigma$-model average
in the presence of an appropriate cosmological constant $\mu$
deformation on the world-sheet\footnote{In the case of
a black-hole coset model this operator is a
`modified cosmological constant' involving some mixing
with appropriate ghost fields parametrising the
$SL(2,R)$ string \cite{bershadsky}.}. The important
point for our discussion is the $\Gamma $-factor
$\Gamma (-s)$. For the interesting case of
matter scattering  off a two-dimensional ($s$-wave four-dimensional)
string
black hole, the latter
is excited to a `massive' (topological) string state
\cite{emnwhair} corresponding to a positive integer
value for $s=n^+ \in {\bf Z}^+$ \cite{emndollar}.
In this case, the expression (\ref{Liouvcorr}) needs
regularization. By employing the `fixed area constraint'
\cite{DDK} one can use an integral representation for
$\Gamma (-s)$
\be
\Gamma (-s)=\int dA e^{-A} A^{-s-1}
\label{integralA}
\ee
where $A$ is the covariant area of the world-sheet. In the
case $s=n^+ \in {\bf Z}^+$ one can then employ a
regularization by analytic continuation,
replacing (\ref{integralA}) by a contour integral
as shown in fig. 1 \cite{kogan2,emndollar,emnshort}.
This is a well-known method of regularization
in conventional field theory, where integrals of
form similar to (\ref{integralA}) appear in terms of
Feynman parameters.
\begin{figure}
\vspace{2.0in}
\caption{           - Contour
of integration in the analytically-continued
(regularized) version of $\Gamma (-s)$ for $ s \in Z^+$.
This is known in the literature as the Saalschutz contour,
and has been used in
conventional quantum field theory to relate dimensional
regularization to the Bogoliubov-Parasiuk-Hepp-Zimmermann
renormalization method.}
\end{figure}
\begin{figure}
\vspace{3.0in}
\caption {          - Schematic repesentation
of the evolution of the world-sheet area as the renormalization
group scale moves along the contour of fig. 1.}
\end{figure}
We note that it is the same regularization which was also
used to prove the equivalence
of the Bogolubov-Parasiuk-Hepp-Zimmerman renormalization prescription
to the dimensional regularization of `t Hooft \cite{BPZ}.
One result of such an analytic continuation is the
appearance of imaginary parts in the respective correlation functions,
which in our case are interpreted \cite{kogan2,emndollar,emnshort}
as renormalization group instabilities of the system.
\pr
Interpreting
the latter as an actual time flow,
we then interpret the contour of fig. 1 as implying
evolution of the world-sheet area in both
(negative and positive) directions of time
 (c.f. fig. 2), i.e.
\be
 Infrared ~ fixed ~ point  \rightarrow  Ultraviolet ~ fixed
{}~point \rightarrow
 Infrared ~ fixed ~ point
\label{flow}
\ee
In each half of the world-sheet diagram of fig. 2,
the Zamolodchikov $C$-theorem
tells us that we have an
irreversible Markov process.
According to the analysis of section 2,
the physical system will be time-irreversible if
the
associated $\Lambda$ transformations are not equivalent.
It has been argued in ref. \cite{emntop} that
a highly-symmetrical phase of the two-dimensional
black hole occurs at the infrared fixed point of the
world-sheet renormalization group flow. At that point, the
associated $\sigma$-model is a topological theory
constructed
by twisting \cite{witt,eguchi} an appropriate $N=2$ supersymmetric
black-hole $SL(2,R)$ Wess-Zumino $\sigma$-model.
The singularity of a stringy black hole,
then, describes a topological degree of freedom.
The highly-symmetric phase is interpreted as
the state with the most `appropriate' initial conditions,
whose preparation requires finite entropy.
This in turn implies a `bounce' interpretation
of the renormalization group flow of fig. 2, in which
the infrared fixed point is a `bounce' point,
similar to the corresponding picture
in point-like field theory \cite{coleman}.
Thus,
the ``physical'' flow of time
is taken to be {\it opposite} to the conventional
renormalization group flow, i.e. from the infrared
to the ultraviolet fixed point on the world sheet.
In the next section we shall see this explicitly,
by using world-sheet instanton calculus
to represent, at least qualitatively,
the renormalization
group flow of the effective target-space theory,
providing a concise expression for the effects of the
topological modes that are linked to the tachyon modes by
$W$ symmetries.
\pr
\section{The String
Black Hole Model and its
World-Sheet Instantons}
\pr
In this section we review some calculations we have made
of specific contributions to the ${\nd S}$ matrix due
to topologically non-trivial world-sheet configurations.
The action of $SL(2,R)/U(1)$ coset Wess-Zumino
model \cite{witt}
describing a Euclidean black hole can be written
in the form
\be
S=\frac{k}{4\pi} \int d^2z \frac{1}{1+|w|^2}\partial _\mu {\bar w}
\partial ^\mu w + \dots
\label{threev}
\ee
where the conventional radial
and angular coordinates $(r,\theta)$ are given
by $w=sinh r e^{-i\theta}$ and the target
space $(r,\theta)$ line element is
\be
ds^2=\frac{dwd{\overline w}}{1 + w{\overline w}}=dr^2+tanh^2rd\theta^2
\label{fourv}
\ee
The Euclidean black hole can be written
as a vortex-antivortex pair \cite{emndua}, which is a solution of the
following Green function equations on a spherical world sheet:
\be
\partial _z\partial _{\bar z} X_v =i\pi \frac{q_v}{2}
[\delta (z-z_1)-\delta(z-z_2)]
\label{fivev}
\ee
The world-sheet can also accommodate
monopole-antimonopole pairs \cite{emndua}, which are solutions of:
\be
   \partial _z  \partial _{\bar z} X_m =-\frac{q_m \pi} {2}
[\delta (z-z_1) -\delta (z-z_2)]
\label{sixv}
\ee
These are related to Minkowski black holes
with masses $\propto q_m$. Vortex and monopole configurations
can both be regarded as sine-Gordon deformations
of the effective action for the field $X\equiv \beta^{-\frac{1}{2}}
{\tilde X}$, where $\beta^{-1}$ is an effective
`pseudo-temperature': $\beta =\frac{3}{\pi (C-25)} $ in Liouville
theory. The partition function \cite{ovrut}
\bea
Z&=&\int D{\tilde X} exp(-\beta S_{eff}({\tilde X}) )  \nn \\
\nonumber
\beta S_{eff}&=&  \int d^2 z [ 2\partial {\tilde X}
{\overline \partial } {\tilde X} +  \frac{1}{4\pi }
[ \gamma _v\omega ^{\frac{\alpha}{2}-2}
(2 \sqrt{|g(z)|})^{1-\frac{\alpha}{4}}: cos (\sqrt{2\pi \alpha }
[{\tilde X}(z) + {\tilde X}({\bar z})]):   \\
& +&   (\gamma _v, \alpha,
{\tilde X}(z) + {\tilde X}({\bar z}) )
\rightarrow (
\gamma _m, \alpha ', {\tilde X}(z) - {\tilde X}({\bar z}))]]
\label{sevenv}
\eea
requires for its specification an angular ultraviolet
cut-off $\omega$ on the world-sheet.
Here $\gamma_{v,m}$ are the fugacities for vortices and spikes
respectively, and $\frac{\alpha}{4} $ is the conformal
dimension $\Delta$. This deformed sine-Gordon theory
has a low-temperature
phase in which monopole-antimonopole pairs are bound
in dipoles as irrelevant deformations
with the conformal dimension
\be
      \Delta _m =\frac{\alpha _m}{4}=
      \frac{\pi\beta}{2}q_m^2 > 1
\label{eightv}
\ee
Monopole-antimonopole pairs correspond to the creation
and anihilation of a microscopic black hole
in the space-time foam.
\pr
As shown in ref. \cite{yung},
the $SL(2,R)/U(1)$  Wess-Zumino coset model
describing a Euclidean black hole also has instantons
given by the holomorphic function
\be
 w(z)=\frac{\rho}{z-z_0}
\label{ninev}
\ee
with topological charge
\be
 Q=\frac{1}{\pi}\int d^2z \frac{1}{1+|w|^2}[{\overline \partial}
{\overline w}\partial      w  - h.c. ]
 =-2 ln(a) + ~const
\label{tenv}
\ee
where $a$ is an ultraviolet cut-off discussed later.
The instanton action on the world-sheet  also depends
logarithmically on the ultraviolet cut-off. As in the case of
the more familiar vortex configuration in the Kosterlitz-Thouless
model, this logarithmic divergence does not prevent
the instanton from having important dynamical effects.
The instanton-anti-instanton vertices take the form \cite{yung}
\be
V_{I{\overline I}}\propto -\frac{d}{2\pi}
\int d^2z \frac{d^2\rho}{|\rho|^4}
e^{-S_0} (e^{(\frac{k[\rho\partial {\overline w} + h.c. + \dots ]}
{f(|w|)}}+ e^{(\frac{k[\rho \partial w + h.c. + \dots]}{f(|w|)}})
\label{elevenv}
\ee
introducing a new term into the effective
action. Making a derivative expansion
of the instanton vertex and taking the large-$k$
limit, i.e. restricting our attention
to instanton sizes $\rho \simeq a$, this new term
has the same form as the kinetic term in (\ref{threev}),
and hence corresponds to a renormalization of
the effective level parameter in the
large $k$ limit:
\be
 k \rightarrow k - 2\pi k^2 d'
\qquad : \qquad
 d' \equiv d\int \frac{d|\rho|}{|\rho|^3}
\frac{a^{2}}{[(\rho/a)^2 + 1]^{\frac{k}{2}}}
\label{twelvev}
\ee
If other perturbations are ignored,
the instantons are irrelevant deformations
and conformal invariance is maintained.
However, in the presence of ``tachyon'' deformations,
$T_0 \int d^2z {\cal F}_{-\frac{1}{2}, 0,0}^{c,c}$
in the $SL(2,R)$ notation of ref. \cite{chaudh},
there are extra logarithmic infinities
in the shift (\ref{twelvev}), that are visible in the dilute
gas and weak-``tachyon''-field approximations.
In this case, there
is a contribution to the effective action of the form
\be
T_0\int d^2z d^2z'<{\cal F}_{-\frac{1}{2}, 0, 0}^{c,c}
(z,{\bar z}) V_{I{\overline I}} (z',{\bar z}')>
\label{tachdeform}
\ee
Using the
explicit form of the ``tachyon''
vertex ${\cal F}$ (\ref{tachyon},\ref{wseven})
given by $SL(2,R)$ symmetry
\cite{chaudh}, it is straightforward
to isolate a logarithmically-infinite contribution
to the kinetic term in (\ref{threev}), associated
with infrared infinities on the world-sheet
expressible in terms of the world-sheet area $\Omega /a^2$
\cite{emndollar,emnnew},
\bea
        gT_0 \int d^2z' \int
\frac{d\rho}{\rho} (\frac{a^2}{a^2 + \rho^2})^{\frac{k}{2}}
\int d^2 z \frac{1}{|z-z'|^2}
\frac{1}{1 + |w|^2}
\partial _{z'} w(z')
\partial _{\bar z'} {\overline w}(z') + \dots \nn \\
\propto gT_0 ln \frac{\Omega}{a^2} \int d^2z'
\frac{1}{1 + |w|^2}
\partial _{z'} w(z')
\partial _{\bar z'} {\overline w}(z')
\label{analyticexp}
\eea
Such covariant-scale-dependent contributions
can be attributed to Liouville field dynamics, through
the ``fixed-area constraint'' in the Liouville path
integral \cite{DDK,kutasov}. The
zero-mode part
can be absorbed in a
scale-dependent shift of $k$\cite{emndollar},
which for large $k >>1 $ may be assumed to exponentiate:
\be
k_R\propto (\frac{\Omega}{a^2})^{(const). \beta ^I T_0 }
\label{thirteenv}
\ee
where $\beta ^I$ is the instanton $\beta$-function \cite{yung}.
In ref. \cite{emndollar} we gave general arguments
and verified to lowest order that instantons represent
massive mode effects, enabling us to identify
$\beta^I =-\beta ^T $, where
$\beta^T$ is
the renormalization-group $\beta$-function of a
matter deformation of the black hole\footnote{Notice that this
implies that the matter $\beta$-function has to be computed
in a non-perturbative way, which is consistent with the
exact conformal field theory analysis of ref. \cite{chaudh}.}.
Notice that in (\ref{thirteenv}) both the
infrared and the ultraviolet cut-off scales enter.
In the following we shall not distinguish between
infrared and ultraviolet cut-offs. The physical
scale of the system, which varies along a
renormalization group trajectory, is the dimensionless
ratio of the two, which is identified with the Liouville
field.
\pr
The change in $k$ and the associated change in the
central charge $c=\frac{3k}{k-2}-1$
and the black-hole mass
$M_{bh} \propto (k-2)^{-\frac{1}{2}}$
do not conflict with any general theorems.
An analogous instanton renormalization
of $\theta$ (c.f. $k$) has been demonstrated \cite{pruisk}
in related $\sigma$-models
that describe the Integer Quantum Hall Effect (IQHE), discussed further
in section 7. Instanton renormalization of $k$ can also be seen
in the Minkowski black hole model of ref. \cite{witt}, defined
on a non-compact manifold
$SL(2,R)/O(1,1)$\footnote{The $\sigma$-model action
of such a theory contains \cite{yung3},
in addition to the action (\ref{threev}),
a total-derivative $\theta$-term
which
can be thought of as a deformation of the
black hole by an ``antisymmetric tensor''
background, which in two dimensions
is a discrete mode as a result of the abelian gauge symmetry.
Its Euclideanized version
has also instanton solutions of the form (\ref{ninev}),
but with {\it finite} action,
which
induce
``Liouville''-time-dependent shifts
to $k$, prior to matter couplings.}.
In our case, as we have seen, the instantons
reflect
a shift of the central charge between the matter
and background sectors of a combined matter $+$ black hole theory,
in which the total
central charge is unchanged.
They correspond to
a combination of world-sheet deformation operators in
the Wess-Zumino model \cite{witt}:
the exactly marginal
operator $L_0^2 {\overline L}_0^2 $ and the irrelevant part
of the exactly-marginal deformation $L_0^1 {\overline L}_0^1$,
which involves an infinite sum
of massive string operators \cite{chaudh}, as we saw in section
4 [see equations (\ref{margintax}),(\ref{chlyk})].
The fact that the $L_0^2{\overline L}_0^2 $ operator rescales
the target-space metric by an overall constant, implies
that
such perturbations have
the same effect as the instanton.
Thus the
instanton represents the effects of massive string
modes that are related to each other and to massless excitations
by a $W$ symmetry.
Matrix elements of the full exactly
marginal light matter $+$ instanton operator have
no dependence on the ultraviolet cut-off $a$, but the separate
matter and instanton parts do depend on $a$, as we have
seen above.
\pr
Since instantons rescale the target-space metric
and the black hole mass, they may also be used
to represent black hole decay.
This is higher-genus effect in string theory \cite{emndec},
so one should expect that instantons could reflect the contributions
of higher genera. This expectation is indeed supported by
an explicit computation of instanton effects in a dilute-gas
approximation in the presence of dilatons.
It is well known \cite{fischler} in
$\sigma$-model perturbation theory theory that
modular infinities of resummed
world-sheet surfaces may be absorbed as an effective renormalization
of lower-genus Riemann surfaces,
above and beyond the local renormalization
effects at fixed genus. For example, modular infinities
for the sphere and torus may be summarized
by a logarithmically-divergent contribution to the dilaton background
$\Phi$ \cite{cohen}
\be
\Phi = \Phi_R + (\frac{d-26}{6} + \frac{d+2}{4})ln|\Lambda /a| +
O[\frac{1}{a^2}]
\label{torus}
\ee
for a string propagating in a $d$-dimensional target space. The
first term $(d-26)/6$
in the coefficient of the logarithmic divergence
is
the world-sheet sphere
contribution
to the conformal
anomaly, whilst the second term is due to the torus.
It has the effect of increasing the effective
central charge, driving a critical model superciritical,
where it is subject to the renormalization group
instability discussed in section 4.
There is a
 similar instanton
 effect in the string black hole model (\ref{threev}) when
one considers the dilaton term, which takes the
form
\be
exp(\frac{1}{8\pi}\int d^2z R^{(2)}[ ln(1 + |w|^2) + \dots ]
\label{dila}
\ee
where $R^{(2)}$ is the world sheet curvature. It is convenient
for technical reasons to use conformal invariance to
concentrate the world-sheet curvature at a point $z^*$
\be
R^{(2)}=4\pi\chi \delta^{(2)}(z - z^*)
\label{deltafun}
\ee
where $\chi =2$
for a world
sheet with spherical topology. Taking the limit
$z^* \rightarrow \infty$, and integrating over
the location $z_0$ of the instanton
one finds the following contribution to the
partition function
\be
\int \frac{d^2\rho}{\rho ^4} (\frac{a^2}{\rho^2})^{\frac{k}{2}}
(\frac{\rho ^2}{\Lambda ^2})
\label{instcon}
\ee
where $\Lambda$ is the infrared cutoff. Since the
leading instanton contributions to the path-integral come from
$\rho =O(a)$, the ultraviolet cut-off,
we find that instantons contribute
\be
\int d^2z R^{(2)}ln|\Lambda / a|
\label{finalcon}
\ee
to the world-sheet effective action. This
reproduces the above mention higher-genus effects
(\ref{torus}). Such shifts in the dilaton are essential
for the consistency of the full string theory within
the interpretation of target time as a local
renormalization scale.
\pr
We conclude this section by discussing the r\^ole of
instantons in the renormalization group flow, in particular
to justify the bounce picture discussed in section 4.
Close to the infrared fixed point, the system is
believed to be topological {\it both} on the world sheet
and in target space-time. A topological $\sigma$-model
is described by an appropriate twist of the ${N = 2}$
supersymmetric ${SL(2,R)/U(1)}$ Wess-Zumino model,
under which the fermions of the ${N = 2}$ model
become ghosts \cite{witt,eguchi,emntop}. The
topological model possesses an enhanced
symmetry which includes a bosonic
${W_{\infty} \otimes W_{\infty}}$ symmetry.
The breaking of ${W_{\infty} \otimes W_{\infty}}$
down to a single $W_{\infty}$ generates space-time
dynamically \cite{emntop}. This breaking of twisted supersymmetry
(or topological BRST symmetry) arises from
instanton effects \cite{yung3}, associated with the
appearance in the presence of such configurations
of logarithmic divergences in correlation
functions, whose
form has been computed in the dilute-gas approximation.
It is essential for the discussion of this effect to
include a constant $u$ in the definition of the
instanton field \cite{yung3}
\be
w=u + \frac{\rho}{z-z_0}
\label{uinst}
\ee
The r\^ ole of the field $u$ is similar to that of the
Higgs field in supersymmetric models
\cite{supra} where the instantons break
supersymmetry dynamically. In that case the field $u$
labels vacua of the theory.
In the limit where the infrared cutoff $\Lambda$
is large: ${\Lambda/a >> u}$, in which case the
relevant correlator has a double-logarithmic divergence
\cite{yung3}:
\be
<O (x_1)O(x_2)>=-8\pi^2 g^I ln\frac{|x_1 -x _2|}{a}ln(\Lambda
|x_1 - x_2|)
 + \dots
\label{ione}
\ee
where the $\dots $ denote subleading terms in the
infrared limit. The dependence of the correlator
(\ref{ione}) on the distance between points
on the world sheet indicates that the world-sheet
topological
symmetry is broken, with the appearance of a metric.
On the other hand, the same correlator
vanishes in the ultraviolet limit ${\Lambda \simeq a}$
and the world-sheet
topological symmetry is restored.
The ultraviolet fixed point of the flow
is a stable conformally-invariant background
for the ${c = 1}$ string, which can be
regarded as an appropriate mixing of the
$SL(2,R)/U(1)$ coset with ghost fields \cite{witt2,mukhi}.
In our picture, instantons provide a
qualitative description of this mixing.
\pr
This picture is supported by a computation of
the vacuum energy associated with
instanton-anti-instanton configurations in the
toy topological model described above. Identifying
thevacuum energy with the one-point function
of the anti-instanton vertex in an instanton
background:
\be
E_{vac}^{I {\overline I}}   =   <V_{{\overline I}}>_I=
-g^{{\overline I}}\partial _{x_1} < O(x_1) O(x_2) >|_{x_1 \rightarrow
x_2}
\label{itwo}
\ee
and recalling that the dominant anti-instanton
configurations have sizes ${\rho \simeq a}$, we can use
(\ref{ione}) to estimate that in the infrared limit
when ${\Lambda/a >> u}$
\be
E_{vac} =16\pi^2 g^I g^{{\overline I}}
\frac{V^{(2)}}{a^2}[ ln(\Lambda /a) + O(1)]
\label{ithree}
\ee
where $V^{(2)}$ is the world-sheet volume.
This logarithmic divergence will be removed in the
full theory when all topological string modes are
taken into account, but the vacuum energy of the effective
theory of locally-measurable observables will be non-zero.
On the other hand, the limit ${u \rightarrow \infty}$
yields zero vacuum energy \cite{yung3}, as
expected for a theory with unbroken
BRST symmetry. According to our previous discussion
this limit coincides with the critical $c=1$ string.
Indeed, it has been argued in \cite{mukhi}
that the $c=1$ string theory could be topological.
In this case, the topological
nature pertains {\it only}
to the world-sheet, the target space theory
being described by
a flat spacetime, over which the `tachyon' matter
field propagates \cite{mukhi}\footnote{The topological
world-sheet character allows for a formal
resummation
of the world-sheet genera in the $c=1$ string case \cite{mukhi},
and should be contrasted
to the situation at the infrared fixed point, where
both the world-sheet and the target space theories are argued to be
topological \cite{emntop}.}.
Thus the ${c = 1}$ model should be
regarded as the stable ground state, to which the false
string vacua with broken BRST symmetry flow,
justifying the bounce interpretation given
in section 4, according to which
temporal flow is opposite to the renormalization flow,
i.e., from the infrared fixed point to the ultraviolet.
\pr
\section{Valley Contributions to the $\nd{S}$ Matrix }
\pr
We now consider the contributions of monopoles
and instantons to $\nd{S}$ matrix elements
giving transitions between a generic initial-state
density matrix $\rho^{A(in)}_B$      and
final-state density matrix $\rho^{C(out)}_D$. This is
described by an absorptive part of a world-sheet
correlation function
\bea
&~&\sum _{X_{out}}~_{in}<A|D,X>_{out}~_{out}<X,C|B>_{in} =  \nn \\
&=&\sum _{X_{out}}~_{in}<0|T(\phi(z_A)\phi(z_D))|X>_{out}
{}~_{out}<X|{\overline T}(\phi(z_C)\phi(z_B))|0>_{in} = \nn \\
&=&~_{in}<0|T(\phi(z_A)\phi(z_D))
{\overline T}(\phi(z_C)\phi(z_B)|0>_{in}
\label{fourteenv}
\eea
Here we have used the optical theorem \cite{mueller}
on the world sheet,
which is valid because conventional quantum field theory,
and indeed quantum mechanics, remain valid on the world sheet,
to replace the sum over unseen states
$X$ by unity. Next, we use dilute-gas approximations
to estimate the leading monopole-antimonopole
and instanton-anti-instanton contributions to the absorptive part
(\ref{fourteenv}). We expect these to be dominated \cite{khoze}
by valley configurations in the Euclidean functional
integral, so that in a semi-classical approximation
\be
\nd{S} \propto Abs\int D\phi _c exp(-S_v(\phi _c))F_{kin}
\label{fifteenv}
\ee
where the integral is over the collective coordinates $\phi _c$
of the valley, whose action is $S_v(\phi _c)$. The function
$F_{kin}$ depends on kinematic factors, taking generically
the form
\be
F_{kin}=exp(E\Delta R)
\label{sixteenv}
\ee
in the case of a four-point function for large $E\Delta R$,
where $E$ is the centre-of-mass energy and $\Delta R$ is the
valley separation parameter. This enables us to make a saddle-point
approximation to the integral (\ref{fifteenv}), which we then
continue back to Minkowski space.
\pr
Valley trajectories $\psi _v$ have a homotopic parameter
$\mu$ and obey an equation of the form
\be
\frac{\partial S_0}{\partial \psi}|_{\psi =\psi_v}= W_\psi (\mu)
\frac{\partial \psi _v}{\partial \mu}
\label{seventeenv}
\ee
where $W_{\psi} (\mu) $ is a weight function that
is positive definite
and decays rapidly at large distances \cite{yung1,khoze}. We
adapt techniques used in the $O(3)$ $\sigma$-model \cite{dorinst,BW}
to find the
monopole-antimonopole and instanton-anti-instanton valleys in a
reduced version of the $SL(2,R)/U(1)$ model. The separations
of the topological defects and anti-defects are well-defined in
the presence of conformal symmetry breaking, which is provided
in our case by the dilaton field \cite{emndollar}. Valleys can be
found by using the analogy \cite{dorinst}
between $\mu$ and a `time' variable
for defect-anti-defect scattering. We do not discuss here
the details of their construction, but record the results.
\pr
The monopole-antimonopole valley function,
expressed in terms of the original world-sheet
variables, reads
\be
 w(z,{\bar z})=\frac{(v-1/v){\bar z}}{1 + |z|^2}
\label{cocentric}
\ee
where $v$ denotes the separation in the $\sigma$-model
framework. Eq. (\ref{cocentric}) represents a
concentric valley, which can then be mapped into an
ordinary valley by applying appropriate conformal
transformations. The function (\ref{cocentric})
interpolates between a far-separated
monopole-antimonopole pair ($v \rightarrow \infty$)
and the trivial vacuum ($v=1$).
For large but finite
separations the corresponding
valley action leads to the action of a monopole-antimonopole
pair interacting via dipole interactions.
The action of the monopole-antimonopole valley
depends on the angular ultraviolet cut-off $w$ introduced
in section 5:
\be
S_m=8\pi q^2 ln (2)\sqrt{2}e^\gamma
+ 2\pi q^2 ln\frac{2R}{\omega  } +
2\pi q^2 ln[\frac{|z_1-z_2|}{(4R^2 + |z_1|^2     )^{\frac{1}{2}}}
\frac{4}{(4R^2 + |z_2 |^2     )^{\frac{1}{2}}}]
\label{eighteenv}
\ee
for a monopole and antimonopole pair of equal
and opposite charges $q$, which we treat as a
collective coordinate over which we must integrate,
where $\gamma$ is Euler's constant, the second term
in (\ref{eighteenv}) is a logarithmically-divergent
self-energy term on a spherical world sheet of
radius $R$, and the last term in
(\ref{eighteenv}) is a dipole interaction energy.
For finite separations $0 < |z_1 -z_2 | < \infty $
and very small
world-sheets $R=O(a \rightarrow 0)$, the action (\ref{eighteenv})
yields
\be
S_m=2\pi q^2 ln  \frac{a}{\omega  } + finite~parts
\label{nineteenv}
\ee
where the ultraviolet cut-off dependence is apparent.
\pr
To construct the instanton valley, we notice that
in the reduced model used for the construction
of the monopole valley (\ref{cocentric})
the solution for an instanton-anti-instanton pair
is derived from the corresponding monopole case
via a conformal transformation in the $(\mu, ln|z|)$-plane.
In ref. \cite{emndollar} we give arguments why this construction
is true for finite separations as well, thereby leading
to an expression of the instanton valley as an (approximate)
conformal transform of the monopole valley (\ref{cocentric}).
The action of the instanton-anti-instanton
valley in the large-separation limit of the
dilute-gas approximation is
\be
S_{I{\overline I}}=kln(1 +|\rho |^2/a^2) +
O(\frac{\rho {\overline \rho}}{(\Delta R)^2)})
\label{twentyv}
\ee
where $\Delta R$ is the separation of an instanton of
size $\rho$ and an anti-instanton of size ${\overline \rho}$,
and we find a dependence on the ultraviolet
cut-off $a$.
The actions (\ref{nineteenv},\ref{twentyv}) substituted
into the general expression (\ref{fifteenv}) make non-trivial
contributions to the $\nd{S}$ matrix that do not
factorize as a product of $S$ and $S^\dagger$ matrix
elements, as we shall now see.
\pr
In the dilute-gas approximation introduced
in section 5, the topologically trivial zero
monopole-antimonopole, zero instanton-anti-instanton
sector in the unitary sum in (\ref{fourteenv}) provides
the usual $S$-matrix description
of scattering in a fixed background, with no
back reaction of the light matter on the metric.
This result is well-known in the
conformal field theory approach to critical
string theory, and is discussed explicitly
in the present context in section 6 of ref. \cite{emndollar}.
This $S$-matrix contribution corresponds to the usual
Hamiltonian description of quantum mechanics, via the
representation $S=1 +i T$ : $T=\int _{-\infty}^{\infty}
dt H(t)$. Any topologically non-trivial contribution
to the unitarity sum in (\ref{fourteenv}) goes beyond
the usual treatment of conformal field theory in critical strings,
and
makes a contribution
to the non-factorization of the $\nd{S}$-matrix : $\nd{S}=S S^\dagger
+ \dots$. Two such contributions that we have identified above
come from the monopole-antimonopole and instanton-anti-instanton
sectors discussed above, which we expect to be dominated by the
valley actions (\ref{eighteenv}) and (\ref{twentyv}) respectively.
\pr
The dependence of the monopole-antimonopole valley
action (\ref{eighteenv}) on the ultraviolet
cutoff $\omega$,
which we identify with the target-space time $t=-ln\omega$,
and of the instanton-anti-instanton valley action (\ref{twentyv})
on the local scale-dependent level parameter $k$ (\ref{twelvev},
\ref{thirteenv}) where $t=-ln a$, tell us that both
valleys contribute to the non-Hamiltonian term in
the modified quantum Liouville equation (\ref{two}),
that are proportional to the anomalous dimensions
${(\Delta_m -1)}$ (\ref{eightv}) of an irrelevant dipole-like
monopole-antimonopole pair
and ${\gamma _0}$
(\ref{thirteenv}) of a matter deformation
respectively.
Integrating up the corresponding modified quantum
Liouville equation (\ref{two}), we find that a generic
${\nd S}$ matrix element, defined at finite time $t$
by ${\rho (t)}$ = ${\nd S (t)}$ ${\rho (0)}$,
contains a factor
\be
 \nd{S} \simeq e^{-2(\Delta _m -1) t + \dots }
\label{twentythreev}
\ee
associated with an irrelevant dipole-like monopole-antimonopole
pair, and
\be
\nd{S} \simeq e^{-2\gamma _0 t + \dots }
\label{twentyfourv}
\ee
from the instanton-anti-instanton valley. Both
of these time-dependences apply in limits of far-separated
defect and anti-defect.
\pr
Before discussing the r\^ole
of topologically non-trivial
world-sheet configurations in
the suppression of coherence at large times,
we review a similar phenomenon
in Hall conductors, namely the
suppression of spatial correlations by
``de-phasons'' \cite{libby}. As we discuss in more detail
in section 7, the two-dimensional
black-hole model is analogous
to a fractional Hall conductor \cite{emndua},
with the Wess-Zumino level parameter $k$ corresponding
to the transverse conductivity.
Hall systems generally are described
by appropriate $\sigma$-models with Wess-Zumino $\theta$-terms,
defined on the two-dimensional space of electron motion
\cite{libby}. The fields of such
$\sigma$-models, which are space-time coordinates in the
black-hole case, correspond to electrons
propagating in the plane,
with the transverse and longitudinal
conductivities $\sigma _{\mu\nu}$
corresponding to background fields in the black-hole case.
The Wess-Zumino terms are associated with instantons
that renormalize non-perturbatively these conductivities\cite{pruisk} :
\be
   \beta _{\mu\nu}=\frac{d\sigma _{\mu\nu}}{d ln L}\ne 0
\label{pruiskenbeta}
\ee
where $ L $ is an infrared  cut-off on the instanton size
that serves as a renormalization group scale\cite{pruisk}.
The effects of these instantons are  seen clearly  in the case of a
$\sigma$-model defined on a compact manifold
$U(m+n)/U(m)\otimes U(n)$,
where $m$,$n$ are electron field replicas
with the physical case corresponding to $m$ and $n \rightarrow 0$.
This limit is equivalent to the corresponding
limit of the non-compact models $U(m,n)/U(m)\otimes U(n)$,
where the corresponding instantons have infinite action
when $m,n \ne 0$, and
might  naively be thought unimportant.
However, the physical limit of $m$ and $n \rightarrow 0$ is
sensitive to instanton effects. Another example of
the importance of infinite-action topological
solutions is provided by three-dimensional
anyonic Chern-Simons theories \cite{lee}, which
are relevant to the fractional quantum Hall effect (FQHE).
\pr
We believe that localization in Hall systems
is directly related to our problem of quantum coherence.
In the IQHE model of ref. \cite{libby}, impurities
are responsible for the localization
of the electron wave function in the plane.
The localization is achieved formally
by representing collectively the effects of impurities
on elelctron trajectories via extended, static
scattering centres termed ``de-phasons'',
which trap the
electron waves into localized
states with sizes $O(1/\sqrt{\rho})$,
where $\rho$ is the de-phason density.
As a result, the electron correlation functions
are suppressed at large spatial separations:
\be
   \propto exp[-(x-y)^2 \rho]
\label{dephason}
\ee
at zero magnetic field ($\theta = 0$).
As the magnetic field is varied so that
the transverse conductivity becomes a  half-integer
(in units of $e^2/h$),
corresponding to a discrete value of the
instanton angle  $\theta = \pi$, the property
of the
de-phasons to destroy phase coherence between
the advanced and the retarded electron propagators
is lost. Quantitatively \cite{libby}, the expectation value of an
electron loop that encircles a de-phason,
in the presence of a magnetic field, is
proportional to
$e^{-(x-y)^2\rho cos\frac{\theta}{2}}$.
Thus, for $\theta = \pi$ the
``effective de-phason density '' $\rho cos(\theta /2)$
vanishes, and
the electrons delocalize implying
a non-zero longitudinal
conductivity.
This delocalization property
is responsible for the transition
between two adjacent plateaux of the
transverse conductivity in the Hall
conductivity diagram \cite{libby}.
These ideas can be
extended to the FQHE \cite{lutken} via the
three-dimensional anyonic Chern-Simons
theories mentioned above, which are closer
to our black-hole interests\footnote{In fact,
it appears to be the zero-field Hall
effect that describes physics at the space-time singularity
\cite{emndua}.}.
In our case, the massive modes of the $SL(2,R)/U(1)$
black-hole model are the analogues of the
de-phasons.
As discussed earlier in this section, the instantons
in this model renormalize the Wess-Zumino
level-parameter $k$ (c.f. $\theta$),
changing the mass and size of the black hole.
The delocalized phase at $\theta =\pi$
may be identified with the ``topological''
phase
at the space-time singularity \cite{emntop}, which is an infrared
fixed point. The propagating
``tachyon''  mixes in this limit, as
we have discussed above, with the
delocalized topological modes of the string that are
analogous to the de-phasons.
The localization properties are consistent
with shrinking of the world-sheet as
one approaches the ultraviolet fixed point that
corresponds to a flat target space-time where the
tachyons are normal localized fields that do not mix with
topological modes.
\pr
Our formalism for the
time evolution of the density matrix is analogous
to the Drude model of quantum
friction \cite{vernon,cald}, with the
massive string modes playing the r\^oles of
`environmental oscillators'. In the language of
world-sheet $\sigma$-model couplings $\{ g \}$,
the reduced density matrix of the observable states is given, relative
to that evaluated in conventional Schr\"odinger quantum mechanics,
by an expression of the form
\be
 \rho (g, g',  t  ) / \rho_S (g, g',  t  )
\simeq
e^{-\eta \int_0^{t} d\tau \int_{\tau ' \simeq \tau }
d\tau '
 \beta ^i G_{ij} \beta ^j } \simeq
e^{ - Dt ({\bf g}  - {\bf  g'} )^2 + \dots }
\label{twentyonev}
\ee
where $\eta$ is a calculable proportionality coefficient,
and $G_{ij}$ is the Zamolodchikov metric \cite{zam}
in the space
of couplings. In string theory, the identification of the target-space
action with  the Zamolodchikov $C$-function $C(\{g \})$ \cite{zam}
enables the Drude exponent to be written in the form
\be
\beta^iG_{ij}\beta^j = \partial _t C(\{g\})
\label{twentytwov}
\ee
which also determines the rate of increase
of entropy
\be
{\dot S}=\beta^i G_{ij}\beta^j S
\label{twentytwob}
\ee
In the string analogue (\ref{twentytwov}) of the Drude model
(\ref{twentyonev}) the r\^ole of the coordinates in (real) space
is played by the $\sigma$-model couplings $g^i$ that are
target-space background fields. Relevant for us
is the tachyon field $T(X)$, leading us to interpret
$(g-g')^2$ in (\ref{twentyonev}) as \cite{emndollar}
\be
(g-g')^2=(T-T')^2 \simeq (\nabla T)^2 (X-X')^2
\label{new}
\ee
for small target separations $(X,X')$. Equation (\ref{new})
substituted into (\ref{twentyonev}) gives us a
suppression very similar to the IQHE case (\ref{dephason}).
\pr
The effect of the time-dependences (\ref{twentythreev},\ref{twentyfourv},
\ref{twentyonev},\ref{new}) is
to suppress off-diagonal elements
in the target configuration space
representation \cite{emohn} of the out-state
density matrix :
\be
\rho _{out} (x,x') ={\hat \rho }(x)\delta (x-x')
\label{twentyfivev}
\ee
This behaviour can be understood intuitively \cite{emndollar,emnshort}
as being related to the apparent shrinking of the string
world sheet in target space, which destroys interferences
between strings localized at different points in target
configuration space, c.f.
the de-phasons in the Hall model \cite{libby}. This
behaviour is generic for
string contributions to the space-time foam, which
make the theory supercritical locally, inducing renormalization
group (target time) flow. The two specific
contributions (\ref{twentythreev},\ref{twentyfourv})
to this suppression (\ref{twentyfivev}) of space-time
coherence that we have identified
in this paper correspond (\ref{twentythreev})
to microscopic black
hole formation and (\ref{twentyfourv}) to the back-reaction
of matter on a microscopic black hole, entailing in each case
information loss across an event horizon.
\pr
\section{$W$ Symmetries and Non-Hamiltonian Time Evolution}
\pr
At various places in earlier sections, we have mentioned the
infinite-dimensional $W$ algebra underlying the string black
hole model, and its r\^ole in interrelating different
solitonic states in the spectrum of the model. In earlier
papers \cite{emn1} we have stressed the r\^ole of $W$ symmetry in
preserving quantum coherence when back-reaction is neglected.
Subsequently, we have argued that the coherence of the
effective theory of light states is suppressed when
back-reaction is taken into account, in particular via
the monopole-antimonopole and instanton-anti-instanton
contributions discussed in the previous section.
In this section we discuss the $W$-transformation properties
of these explicit contributions to ${\nd S}$ and
${\nd \delta H}$, and relate
them to the formalisms of refs \cite{misra} and \cite{santilli}.
In this way, we link explicitly the loss of coherence to
the leakage of $W$ quantum numbers.
\pr
As a warm-up, we first present an analogous phenomenon in the
theory of the Quantum Hall Effect \cite{eliashvilli}. The
ground state of an
integer Quantum Hall conductor with filling fraction
${\nu = 1}$ is represented by a
non-singular wave function ${\Psi _0 (z_1 , z_2 , ... , z_N )}$
for a system of non-interacting electrons.
On the other hand, the ground state of a fractional
Quantum Hall conductor is singular:
\be
\Psi (z_1 , z_2 , ... , z_N ) = S_p \Psi _0 (z_1 , z_2, ... , z_N )
\label{wone}
\ee
where $p$ is related to the filling fraction $\nu$ by
${\nu = 1/(2p + 1)}$ and
\be
S_p = \Pi_{k<l} (z_k - z_l )^{2p}
\label{wtwo}
\ee
The prefactor $S_p$ can be regarded as the matrix element
of an operator that creates monopoles and vortices on the
world sheet:
\be
 S_p \propto exp(\sum_{k < l}[ 2pRe
ln(z_k - z_l)
+ i2pImln(z_k - z_l)])
\label{wthree}
\ee
This representation realizes the plasma picture of
\cite{laughlin}, and is in direct analogy with the
representation of Minkowski and Euclidean black holes
as monopoles and vortices on the world sheet which we
introduced in section 5.
\pr
The important aspect of this analogy for the purposes
of this section is the observation that the relation
(\ref{wone}) between the IQHE and FQHE ground states
can be regarded as an invertible but non-unitary
relation between the corresponding density matrices:
\be
\rho_p = S_p {S_p}^\dagger \rho_0
\label{wfour}
\ee
The relationship between the
density matrices $\rho_0$ and $\rho_p$ corresponds to that between
the $\rho$ and ${\tilde \rho}$ of \cite{misra}, as we shall
see in more detail shortly.
\pr
It has been pointed out that the IQHE and FQHE systems
both possess an infinite-dimensional $W$ symmetry
associated with the presence of incompressible quantum
electron fluids \cite{trug}, in which the quantum deformation
parameter \cite{bakas} $\frac{1}{k}$
is related to the filling fraction $\nu$
by
\be
\nu = \frac{1}{k}
\label{wfive}
\ee
Thus the operator $S_p$ can be regarded as inducing a
quantum deformation of the $W$ algebra that does not
change the classical $W$ charges. The operator $S_p$
induces $\delta$-function Schwinger terms in the
operator product expansion, associated with the
singularities in $S_p$. They are related to the
corresponding term
in the operator product expansion for two
energy-momentum tensors, which is a
measure of the
central charge $c$ in the Virasoro algebra,
which is related in turn to the level parameter $k$ and
hence to the filling fraction $\nu$:
\be
c=\frac{3k}{k-2}=\frac{3}{1-2\nu}
\label{wsix}
\ee
As we shall see shortly, the world-sheet
monopoles and instantons discussed in section 5
play a similar r\^ole in the black-hole case.
\pr
In the case of a string black hole contribution to
space-time foam, the r\^ole of the
density matrix $\rho$ of \cite{misra} is played
by the exact density matrix for external tachyon
fields dressed by higher-level operators as
discussed in sections 4 and 5:
As is well known \cite{verl}, there is a well-defined $S$-matrix
for the scattering of these dressed tachyons
off a string black hole, and
hence a well-defined Hamiltonian $H$, and the
time evolution of the density matrix is given by
the conventional Liouville equation using this
Hamiltonian.
\pr
However, as we have mentioned previously,
a realistic scattering experiment does not
measure the non-local topological solitonic states
created by the operators $L_0^1 {\bar L}_0^1$ etc.,
and hence deals only with bare tachyonic operators $T$.
As we have described in section 5, the scattering
of these bare operators exhibits additional
renormalization scale (i.e., Liouville field $\phi$, i.e.,
time) dependence that cannot be absorbed within
the usual Hamiltonian/$S$-matrix description. It is the
density matrix of this $(T , \phi )$ system that we
interpret as the ${\tilde \rho}$ of \cite{misra}:
\be
{\tilde \rho}
=\frac{e^{-\beta {\cal H}([{\cal F}^{c-c}
_{-\frac{1}{2},0,0}]_{r,\phi})}}
{Tr[e^{-\beta {\cal H}([{\cal F}^{c-c}
_{-\frac{1}{2},0,0}]_{r,\phi})}]}
\label{weight}
\ee
where ${\cal F}^{c-c}_{-\frac{1}{2}, 0, 0}$ is the `tachyon' deformation
defined in (\ref{tachyon}),
and $[ \dots ]_{\phi}$ denotes the appropriate Liouville dressing.
It is because of the extra time dependence mentioned above
that this density matrix ${\tilde \rho}$ obeys the
modified Liouville equation (\ref{two}) derived
in section 3 \cite{emnqm}.
\pr
The monopole-antimonopole pairs discussed in the previous section
contribute to the relation between
$\rho$ (\ref{margintax}) to ${\tilde \rho}$
(\ref{weight}), by representing the creation and annihilation
of black holes, and the instanton-anti-instanton pairs make
a contribution to this relation that is
associated with rescaling of the target-space metric due
to changes in the black hole mass.
According to the analysis of section 2, this
relation is associated with a loss of quantum coherence,
and we have exhibited just such a loss in section 5.

In order to understand the physical origin of this
loss of coherence,
it is instructive to consider the
$W$-transformation properties of these topological
defects on the world sheet. To do this, we express
the string black hole action in terms of
target-space Kruskal-Szekeres coordinates
$u \equiv sinhr e^{it}$, ${\overline u}=sinhr e^{-it}$:
\be
S=\frac{k}{4\pi} \int d^2z \frac{
\partial u   {\overline
\partial } {\overline u} +
\partial {\overline u}   {\overline
\partial } u           }{1 - {\overline u}u }
\label{wnine}
\ee
Monopoles generate transformations of the form
\be
u \rightarrow u e^{i \alpha}, {\bar u} \rightarrow
{\bar u} e^{-i \alpha}
\label{wten}
\ee
It is useful to construct parafermion operators
$\psi_{+,-}$ by attaching Dirac strings to these
monopoles \cite{bakir}:
\bea
\psi _+ & = &
\frac{\partial u}{\sqrt{1 - {\overline u}u }}V_+   \nn \\
\psi _- & = &
\frac{\partial {\overline u}}{\sqrt{1 - {\overline u}u }}V_-   \nn \\
V_{\pm} & = &
exp[\pm \frac{1}{2} \int _C (dz A + d{\bar z} {\overline A} )]
\nn \\
A & = & \frac{u \partial {\overline u} - {\overline u} \partial u }
{1 - {\overline u} u }  \qquad ; \qquad
{\overline A}   =
\frac{{\overline u} \partial u -  u {\overline \partial} {\overline u} }
{1 - {\overline u} u }
\label{weleven}
\eea
The $W$ currents are then given in terms of these
constructs by
\be
W_s =\sum_{k=1}^{s-1} (-1)^{s-k-1} A_k^s \partial ^{k-1}
\psi _+ \partial ^{s-k-1} \psi_-
\label{wtwelve}
\ee
where
$s$ denotes the conformal spin, and
\be
A_k^s =\frac{1}{s-1}
\left(\begin{array}{c}  s-1   \\
k \end{array} \right)
\left(\begin{array}{c}  s-1   \\
s-k \end{array} \right)
\label{AA}
\ee
We note that all these $W_{\infty}$ transformations
on the world sheet are generated by $(1,0)$ or $(0,1)$
currents, so the corresponding stress-tensor
deformations are $(1,1)$, and hence can surely be elevated
to target space \cite{emn1}. This elevation has been worked out in
\cite{bakir}, where it was shown that this elevation
preserves the $W_{\infty}$ structure. One considers
variations of the action $S$ under infinitesimal
transformations $\delta u$, $\delta {\bar u}$:
\be
\delta S =\int (\delta _{\epsilon} ^{(s)} u \frac{\delta S}{\delta u}
+
\delta _{\epsilon} ^{(s)} {\overline u}\frac{\delta S}{\delta {\overline
u}})
\label{wthirteen}
\ee
where
\bea
\frac{\delta S}{\delta u} = \frac{\partial {\overline \partial} u }
{1 - u {\overline u} } + \frac{{\overline u}\partial u {\overline
\partial} u }{(1-u{\overline u})^2}  \nn \\
\frac{\delta S}{\delta {\overline u}} =
\frac{\partial {\overline \partial} {\overline u} }
{1 - u {\overline u} } + \frac{ u\partial {\overline u} {\overline
\partial} {\overline u} }{(1-u{\overline u})^2}
\label{wfourteen}
\eea
Knowing that the $W_{\infty}$ transformations are
generated by chiral fields, we compare (\ref{wthirteen})
with the Ansatz \cite{bakir}
\be
\delta S = \int d^2z \epsilon {\overline \partial} W_s
\label{wfifteen}
\ee
to find the following infinite set of infinitesimal
transformations:
\bea
\delta _\epsilon ^{(s)} u &=& \sum _{i=0}^{s-2} B_i^s
\partial ^i \epsilon \sqrt{1 - u {\bar u}}V_{-}
\partial ^{s-i-2} \psi _+ +   \nn \\
{}~&~ &\sum_{k=1}^{s-1} \sum_{l=0}^{k-2} (-1)^l A_k^s
\left(\begin{array}{c} k-1 \\ l \end{array} \right)u
\partial ^{k-2-l} [\epsilon \partial ^l \psi_-
\partial^{s-k-1} \psi _+ - \nn \\
{}~&~& (-1) ^s \epsilon
\partial ^l \psi _+ \partial ^{s-k-1} \psi _- ]
\label{wsixteen}
\eea
with
\be
B_l^s =\frac{1}{s-1}
\left(\begin{array}{c} 2s-l-2 \\
s \end{array} \right)
\left(\begin{array}{c}  s-1   \\
l \end{array} \right)
\label{bA}
\ee
The coresponding transformation for ${\overline u}$
is obtained by interchanging  $u \leftarrow\rightarrow {\overline u}$
and multiplying by a factor $(-1)^s$.
\pr
It is easy to verify directly that these variations
satisfy the $W_{\infty}$ algebra:
\be
     [\delta _{\epsilon_1}^{s} ,
\delta_{\epsilon_2}^{s'} ] = \delta _{(s'-1)\epsilon_1'\epsilon _2 -
(s-1)\epsilon _1\epsilon_2'}^{s + s' -2}  + \dots
\label{wseventeen}
\ee
where the $\dots$ denote lower-spin terms.
Hence the $W$ charges
\be
Q^s= \int _C dz W^s (z)
\label{weighteen}
\ee
commute, as they constitute an infinite-dimensional
Cartan subalgebra of $W_{\infty}$ \cite{bakas}.
It is easy to check that these charges are non-zero
in general if $u$ and ${\bar u}$ are not holomorphic
functions of the world-sheet coordinates: $u \ne u(z)$,
etc., which is the case for the string black hole
background. Therefore, their values make available
an infinite set of quantum numbers (hair) to
specify the black hole state.
However, the instantons discussed in section 5
have vanishing $W$ quantum numbers, because they are
holomorphic maps $u=u(z)$. One finds that
\be
\delta _{\epsilon}^s u
= \sum _{l=0} ^{s-2} B_l^s \partial ^l
\epsilon \partial ^{s-l-1} u \qquad ; \qquad
\delta _{\epsilon}^s {\overline u} = 0
\label{wnineteen}
\ee
Note that instantons and anti-instantons behave
differently under
the $W_{\infty}$ transformations.
\pr
Now we are ready to discuss in the light of
these results the monopole and
instanton contributions to the loss of coherence.
As has already been mentioned, monopoles make
contributions to the functional integral
representation of ${\tilde \rho}$ that are
identical in form to the prefactor $S_p$ in
the relation (\ref{wfour}) between the IQHE and
FQHE ground states, and play the same r\^ole as
the similarity transformation of \cite{misra}.
We saw in the previous section [see equation
(\ref{eighteenv})]
that
the action of a monopole-antimonopole pair depends
logarithmically on the ultraviolet cutoff, which
translates into a time-dependence, and hence a
contribution to ${\nd \delta H}$ that tends to
suppress quantum coherence. We have seen in this section
that monopoles possess $W$ hair, by virtue of being
non-holomorphic maps of the world sheet into target
space. Thus they carry off $W$ quantum numbers, which
entails information loss and hence
explains the
loss of coherence.
The action of the instanton-anti-instanton valley exhibits
a logarithmic dependence on the ultraviolet cutoff (and
hence time-dependence, a contribution to ${\nd \delta H}$
and a loss of coherence) only when there is a finite
separation between the instanton and anti-instanton.
As we have discussed just above, an isolated instanton is
represented by a holomorphic map from the world sheet
into target space, hence carries no $W$ quantum numbers
and does not contribute to decoherence. However, once
an instanton and anti-instanton are superposed at finite
separation, their configuration can no longer be
represented by such a holomorphic map, and they will
carry $W$ quantum numbers in general, explaining this
loss of coherence.
\pr
Thus we have a complete understanding of the relation
between our non-critical string formalism and the
general similarity transformation theory of
\cite{misra}, including a physical understanding of the
resulting loss of coherence due to string black holes
in terms of a leakage of
$W$ quantum numbers. We expect that a similar
discussion could be given for any other topologically
non-trivial contributions to the space-time foam in
string theory, in which other string symmetries
would play the r\^ole of the black hole's
$W_{\infty}$ algebra, which is but a small part of
the enormous full local symmetry of string.
\pr
\section{CPT Violation}
\pr
It is a basic theorem \cite{cpt} of quantum field theory that $CPT$
should not be violated, as a consequence of locality,
Lorentz invariance and unitarity. String theory is of
course based on local quantum field theory
on the world sheet, but this is not equivalent to
locality in target space-time. Lorentz invariance is
a property of critical string theory, but requires
re-examination in the context of the non-critical
string approach to time that we have espoused here.
Thus, even though we do not challenge the unitarity
of the effective low-energy theory, not all the
conditions needed to derive the $CPT$ theorem are
satisfied in our string framework, and it is
appropriate to re-open the possibility of $CPT$
violation.
\pr
This has often been discussed \cite{wald}
in the general context of
quantum gravity, motivated largely by the
likelihood of non-locality, and also specifically
in the context of string theory \cite{cptwitt,cptstring,sonoda}. In
this latter
case, it was shown that $CPT$ violation was in principle
possible if certain world-sheet charges were not
conserved \cite{cptwitt}, but it was also shown that this could
not occur in a string model with a flat target
space-time \cite{sonoda}.
\pr
We have taken this analysis one step further,
pointing out that space-times that appear
singular from the point of view of conventional
general relativity, such as a black hole,
can be described by topological defects on the world
sheet, such as monopoles or vortices in the black-hole
case \cite{emndua}. Moreover, it is well-known that quantum numbers
carried by fermionic fields may no longer be
conserved in the presence of topological defects,
c.f., monopoles and instantons in conventional
three-dimensional space. To demonstrate that $CPT$
is violated in any specific elementary-particle
system, such as the neutral kaon system, would
require a complete string-derived model in which
the world-sheet fermion contents of all quarks
and leptons were known. Would that we had such a
complete model! However, even in its absence,
we \cite{emncpt,elmn} and Huet and Peskin \cite{hp} have argued
that the likelihood of $CPT$ violation in string
theory makes it worthwhile to re-examine the
traditional phenomenology of the neutral kaon and
other systems, parametrizing possible
$CPT$-violating effects and constraining their
magnitudes. Before reviewing that work here, we
would like to make contact with other work \cite{wald} on
$CPT$ violation in the general framework of
quantum gravity.
\pr
The conventional, strong form of $CPT$ invariance
proved in local quantum field theory implies the
existence of a $CPT$ operator $\Theta$ that
transforms any initial density matrix $\rho_{in}$
into some final-state density matrix
${\rho_{out}}^{\prime}$:
\be
{\rho_{out}}^{\prime} = \Theta \rho_{in}
\label{cptone}
\ee
and correspondingly
\be
{\rho_{in}}^{\prime} = \Theta^{-1} \rho_{out}
\label{cpttwo}
\ee
where $\rho_{out}$ is related to $\rho_{in}$ by the
familiar ${\nd S}$ matrix:
\be
\rho_{out} = {\nd S} \rho_{in}
\label{cptthree}
\ee
and likewise for ${\rho_{out}}^{\prime}$ and
${\rho_{in}}^{\prime}$:
\be
{\rho_{out}}^{\prime} = {\nd S} {\rho_{in}}^{\prime}
\label{cptfour}
\ee
The following relation is a trivial consequence of
equations (\ref{cptone}) to (\ref{cptfour}):
\be
\Theta = {\nd S} {\Theta}^{-1} {\nd S}
\label{cptfive}
\ee
Again, it is a trivial consequence of (\ref{cptfive})
that the ${\nd S}$ matrix must have an inverse :
\be
{\nd S}^{-1} = {\Theta}^{-1} {\nd S} {\Theta}^{-1}
\label{cptsix}
\ee
However, there can be no inverse of the superscattering
matrix ${\nd S}$ in any framework that allows pure states
to evolve into mixed states, as has been argued to be a
general necessity in any quantum theory of gravity, and
as we have found specifically in our non-critical string
approach. Thus, there cannot be a $CPT$ operator with
the properties assumed above, and the $CPT$ theorem
cannot hold in its strong field-theoretical form.
\pr
A weaker form of CPT invariance has been proposed \cite{wald},
according to which the probability
${P(\phi \rightarrow \psi)}$ that a pure initial
state $\phi$ will be observed to become a given pure
final state $\psi$ is equal to the probability for
the $CPT$-reversed process:
\be
P(\psi \rightarrow \phi) = P(\theta^{-1} \phi
\rightarrow \theta \psi)
\label{cptseven}
\ee
This is automatically true in any theory in which
the final-state density matrix is proportional to the unit
matrix, as was the case in our simple two-state model in
section 3. However, we will not address here the interesting
question whether this or some other weak form of $CPT$
invariance holds in string theory. For now, we simply
observe that there is no reason for
the strong $CPT$ theorem to hold in string theory.
To substantiate this claim we review briefly the
arguments presented in ref. \cite{emncpt}
concerning $CPT$-violation in our string model.
Following \cite{cptwitt} we consider the target-space
$CPT$ operator $\Theta $ (\ref{cptone}) as being
derived from an appropriate world-sheet operator $\Theta _w$.
Any state in target-space can be considered as an eigenstate
of the $\sigma$-model Hamiltonian $E$ with eigenvalue $m_i$,
the mass of the corresponding particle, i.e.
\be
E | m_i, Q_i > = m_i | m_i, Q_i >
\label{cptH}
\ee
where $\{ Q_i \}$ denotes a set of conserved
world-sheet charges that can be elevated to target space,
\be
[ E, {\hat Q}_i ] =0 \qquad ; \qquad  {\hat Q}_i | m_i, Q_i >
= Q_i | m_i, Q_i >
\label{cptQH}
\ee
The ${\hat A} $ denote quantum-mechanical
operators on the world-sheet.
World-sheet $CPT$ invariance is guaranteed
if and only if \cite{cptwitt}
\be
[ E, \Theta _w ] = 0 \qquad ;  \qquad {\hat Q}_i \Theta _w
+ \Theta _w {\hat Q}_i= 0
\label{cptQTheta}
\ee
This implies $CPT$-invariance in target space in the following sense
\cite{cptwitt} : the $CPT$ transform of a state of mass $m_i$ and
`charge' $Q_i$, is
$ \Theta _w | m_i, Q_i > $. Using (\ref{cptQTheta}) we can
readily see that it will be an eigenstate of $E$
with mass $m_i$ and `charge' $-Q_i$.
In our case, the existence of valleys of topological
defects on the world-sheet spoils the conservation
of $Q_i$, and thus (\ref{cptQTheta}), as a result
of logarithmic divergences, as we have discussed in sections 5 and 6.
These imply temporal dependences of the `charges' $Q_i$,
and hence their conservation is spoiled.  As a consequence, the
above `proof' of $CPT$ invariance in target space fails.
\pr
Although the above picture is rather heuristic, and much more
work is required to define the elevation process
of the $CPT$ operation from the world sheet to target space
in a mathematical rigorous way,
however it is certainly suggestive
of the kind of $CPT$ violation one should expect in string theory
formulated in highly curved space-time backgrounds.
It is therefore
worthwhile to explore the possibility of its violation,
though we also cannot exclude the possibility that the strong
$CPT$ theorem might not be violated detectably in any given
experiment.
\pr
We now describe briefly the formalism \cite{ehns,emncpt}
for describing the
possible modification of quantum mechanics and violation of
$CPT$ in the neutral kaon system, which is among the most
sensitive microscopic laboratories for studying these
possibilities. In the normal quantum-mechanical
formalism, the time-evolution of a neutral kaon
density matrix is given by
\be
\partial _t \rho = -i (H\rho - \rho H^{\dagger})
\label{cpteight}
\ee
\nk where the Hamiltonian takes the following form
in the ${(K^0 , {\overline K}^0 )}$ basis:
\be
  H = \left( \begin{array}{c}
 (M + \frac{1}{2}\Delta M) - \frac{1}{2}i(\Gamma + \frac{1}{2}
 \Delta \Gamma)
   \qquad  \qquad
   M_{12}^{*} - \frac{1}{2}i\Gamma _{12}^{*} \\
           M_{12}  - \frac{1}{2}i\Gamma _{12}
    \qquad  \qquad
    (M - \frac{1}{2}
    \Delta M)-\frac{1}{2}i(\Gamma
    - \frac{1}{2}
    \Delta \Gamma ) \end{array}\right)
\label{cptnine}
\ee
The non-hermiticity of $H$ reflects the process of
$K$ decay: an initially-pure state evolving
according to (\ref{cpteight}) and (\ref{cptnine})
remains pure.
\pr
In order to discuss the possible modification of this
normal quantum-mechanical evolution, and allow for the
possibility of $CPT$ violation, it is convenient to
rewrite \cite{emncpt}
(\ref{cpteight}) and (\ref{cptnine}) in a
Pauli $\sigma$-matrix basis \cite{ehns}, introducing
components ${\rho}_{\alpha}$ of the density matrix:
\be
\rho = 1/2 \rho_{\alpha} \sigma_{\alpha}
\label{cptten}
\ee
which evolves according to
\be
 \partial _t \rho_\alpha  =
h_{{\alpha}{\beta}}{\rho_{\beta}}
\label{cpteleven}
\ee
with
\be
  h_{\alpha\beta} \equiv \left( \begin{array}{c}
 Imh_0 \qquad Imh_1 \qquad Imh_2 \qquad Imh_3 \\
 Imh_1 \qquad Imh_0 \qquad -Reh_3 \qquad Reh_2 \\
 Imh_2 \qquad Reh_3 \qquad Imh_0 \qquad -Reh_1 \\
 Imh_3 \qquad -Reh_2 \qquad Reh_1 \qquad Imh_0 \end{array}\right)
\label{cptelevenb}
\ee
It is easy to check that at large times $\rho$ takes the form
\be
  \rho \simeq e^{-\Gamma _Lt}
 \left( \begin{array}{c}
 1   \qquad  \epsilon^*            \\
 \epsilon          \qquad |\epsilon |^2 \end{array}\right)
\label{cpttwelve}
\ee
where $\epsilon$ is given by
\be
  \epsilon =\frac{\frac{1}{2}i Im \Gamma _{12} - Im M_{12}}
{\frac{1}{2} \Delta \Gamma - i\Delta M }
\label{cptthirteen}
\ee
in the usual way.
\pr
A modification of quantum mechanics of the form
discussed in section 3 can be introduced by modifying
equation (\ref{cpteleven}) to become
\be
\partial _t \rho_\alpha  = h_{\alpha\beta} \rho_\beta +
\nd{h}_{\alpha\beta}\rho_\beta
\label{cptfourteen}
\ee
The form of ${\nd h}_{\alpha \beta}$ is determined if we
assume probability and energy conservation, as proved
in the string context in section 3, and that the
leading modification conserves strangeness:
\be
  {\nd h}_{\alpha\beta} =\left( \begin{array}{c}
 0  \qquad  0 \qquad 0 \qquad 0 \\
 0  \qquad  0 \qquad 0 \qquad 0 \\
 0  \qquad  0 \qquad -2\alpha  \qquad -2\beta \\
 0  \qquad  0 \qquad -2\beta \qquad -2\gamma \end{array}\right)
\label{cptfifteen}
\ee
It is easy to solve the $4 \times 4$ linear matrix equation
(\ref{cptfourteen}) in the limits of large time:
\be
\rho _L
\propto \left( \begin{array}{c} 1 \qquad  \qquad
\frac{-\frac{1}{2}i  (Im \Gamma _{12} + 2\beta )- Im M_{12} }
{\frac{1}{2} \Delta \Gamma + i \Delta M } \\
\frac{\frac{1}{2}i (Im \Gamma _{12} + 2\beta )- Im M_{12} }
{\frac{1}{2}\Delta \Gamma  - i \Delta M} \qquad \qquad
|\epsilon |^2 + \frac{\gamma}{\Delta \Gamma } -
\frac{4\beta Im M_{12} (\Delta M / \Delta \Gamma ) + \beta ^2 }
{\frac{1}{4} \Delta \Gamma ^2 + \Delta M^2 } \end{array} \right)
\label{cptsixteen}
\ee
and of short time:
\be
 \rho _S
 \propto \left( \begin{array}{c} |\epsilon |^2 +
\frac{\gamma }{|\Delta \Gamma |} -
\frac{-4\beta Im M_{12} (\Delta M/\Delta \Gamma )+ \beta ^2 }
{ \frac{1}{4} \Delta \Gamma ^2 + \Delta M^2 } \qquad
\epsilon - \frac{i\beta}{\frac{\Delta \Gamma} {2} - i \Delta M} \\
\epsilon ^* + \frac{i\beta} {\frac{\Delta \Gamma}{2} +
i \Delta M} \qquad 1 \end{array} \right)
\label{cptseventeen}
\ee
We note that the density matrix (\ref{cptsixteen}) for
$K_L$ is mixed to the extent that the parameters $\beta$
and $\gamma$ are non-zero. It is also easy to check \cite{emncpt}
that
the parameters $\alpha$, $\beta$ and $\gamma$ all violate
$CPT$, in accord with the general argument of \cite{wald},
and consistent with the string analysis mentioned earlier
in this section.
\pr
Experimental observables $O$ can be introduced \cite{ehns,emncpt}
into this
framework as matrices, with their measured values being given by
\be
<O> = Tr(O \rho)
\label{cpteighteen}
\ee
Examples are the $K$ to $2 \pi$ and $3 \pi$ decay
observables
\be
 O_{2\pi} =\left( \begin{array}{c} 0 \qquad 0 \\
0 \qquad 1 \end{array} \right)
\qquad ; \qquad
 O_{3\pi}
 =(0.22)
 \left( \begin{array}{c} 1 \qquad 0 \\
0 \qquad 0 \end{array} \right)
\label{cptnineteen}
\ee
and the semileptonic decay observables
\bea
  O_{\pi ^-l^+ \nu} = \left(\begin{array}{c} 1 \qquad 1 \\
1\qquad 1 \end{array}\right)  \nn \\
  O_{\pi ^+l^-{\overline \nu}}  =\left( \begin{array}{c}
1 \qquad -1 \\
 -1\qquad 1 \end{array} \right)
\label{cpttwenty}
\eea
A quantity of interest is the difference between
the $K_L$ to $2 \pi$ and $K_S$ to $3 \pi$
decay rates \cite{emncpt}:
\be
\delta R \equiv R_{2\pi}-R_{3\pi}= \frac{8\beta}{|\Delta \Gamma |}
 |\epsilon| sin\phi _{\epsilon}
\label{cpttwentyhalf}
\ee
where
$R_{2\pi}^L\equiv Tr(O_{2\pi} \rho_L)$,
and $R_{3\pi}^S \equiv Tr(O_{3\pi}\rho_S)/0.22 $, and the prefactors
are determined by the measured \cite{pdg}
branching ratio for
$K_L \rightarrow 3{\pi}^0$. (Strictly speaking, there should be a
corresponding prefactor of $0.998$ in the formula
(\ref{cptnineteen}) for
the $O_{2{\pi}}$ observable.)
\pr
Using (\ref{cpttwenty}), one can calculate the semileptonic
decay asymmetry \cite{emncpt}
\be
\delta \equiv \frac{\Gamma (\pi^-l^+\nu) - \Gamma (\pi^+ l^-
{\overline \nu }) }{\Gamma (\pi ^- l^+ \nu ) +
\Gamma (\pi ^+ l^- {\overline \nu})}
\label{cpttwentyone}
\ee
in the long- and short-lifetime limits:
\bea
\delta_L  & = &   2Re [\epsilon (1-\frac{i\beta}{Im M_{12}})]   \nn \\
       \delta_S   & = &
2Re [ \epsilon (1 + \frac{i\beta}{Im M_{12}})]
\label{cpttwentytwo}
\eea
The difference between these two values
\be
          \delta\delta \equiv
\delta _L  - \delta _S  = -\frac{8\beta} {|\Delta \Gamma |}
\frac{sin\phi _{\epsilon}}{\sqrt{1 + \tan^2 \phi _{\epsilon}}}
= -\frac{8\beta}
{|\Delta \Gamma |} sin\phi _{\epsilon}
cos\phi _{\epsilon}
\label{cpttwentythree}
\ee
with $tan\phi _\epsilon =(2\Delta M)/\Delta \Gamma $,
is a signature
of $CPT$ violation that can be explored at the
CPLEAR and DA$\phi$NE facilities.
\pr
We have used \cite{emncpt} the latest experimental values of
$R_{2 \pi}$ and $R_{3 \pi}$ to bound $\delta R$,
and the latest experimental values of $\delta_{L,S}$
to bound $\delta \delta$, expressing the results as
contours in the $(\beta , \gamma )$ plane as seen in
figure 3. Also shown there are contours of the usual
$CP$-violating parameter $\epsilon$, which is given in
our case by \cite{emncpt}
\be
|\epsilon| = -\frac{2\beta}{|\Delta \Gamma |}sin\phi _{\epsilon}
+ \sqrt{\frac{4\beta ^2}{|\Delta \Gamma |^2 } - \frac{\gamma}
{|\Delta \Gamma |} + R_{2\pi}^L }
\label{cpttwentyfour}
\ee
\begin{figure}
\vspace{3.5in}
\caption{          - The
$(\beta, \gamma)$ plane on
a logarithmic
scale for $\beta > 0$. We plot
contours of the conventional
$CP$-violating parameter $|\epsilon |$, evaluated from the
$K_L \rightarrow 2\pi $ decay rate. The dashed-double-dotted
band is that allowed
at the one-standard-deviation level by the comparison
between measurements of the $K_L \rightarrow 2\pi $ decay rate
and the $K_L$ semileptonic decay asymmetry $\delta _L $.
The dashed line delineates the boundary
of the region allowed by the
present experimental upper limit on $K_S \rightarrow 3\pi ^0$
decays $(R_{2\pi}^L-R_{3\pi}^S)$
and a solid line delineates the boundary
of the region allowed by
a recent preliminary measurement of the $K_S$
semileptonic decay asymmetry $\delta _S$.
A wavy line bounds approximately the region of $|\beta |$
which may be prohibited by intermediate-time measurements of
$K \rightarrow 2\pi$ decays.}
\end{figure}
\begin{figure}
\vspace{3.5in}
\caption {         -  As in Fig. 3, on a
logarithmic scale for $\beta < 0$.}

\end{figure}

\begin{figure}
\vspace{4.50in}
\caption        {     As in Fig. 3, on a
linear scale for the neighborhood of $\beta=0$.}
\end{figure}
On the basis of this preliminary analysis, it is safe
to conclude that
\be
|\frac{\beta}{\Delta \Gamma}| \lsim ~10^{-4}~to~10^{-3}
\qquad ; \qquad
|\frac{\gamma}{\Delta \Gamma}| \lsim ~10^{-6}~to~10^{-5}
\label{cpttwentyfive}
\ee
In addition to more precise experimental data,
what is also needed is a more complete global fit
to all the available experimental data, including
those at intermediate times, which are essential for
bounding $\alpha$, and may improve our bounds
(\ref{cpttwentyfive}) on $\beta$ and $\gamma$
\cite{elmn,hp}
{}.

\pr
We cannot resist pointing out that the bounds
(\ref{cpttwentyfive}) are quite close to
\be
O(\Lambda_{QCD} / M_P ) m_K \simeq 10^{-19} GeV
\label{cpttwentysix}
\ee
which is perhaps the largest magnitude that any
such $CPT$- and quantum-mechanics-violating
parameters could conceivably have. Since any
such effects are associated with topological
string states that have masses of order $M_P$,
we expect them to be suppressed by some power of
$1/M_P$. This expectation is supported by the
analogy with the Feynman-Vernon model of
quantum friction \cite{vernon}, in which coherence is
suppressed by some power of the unobserved
oscillator mass or frequency. If the $CPT$- and
quantum-mechanics-violating parameters discussed in
this section are suppressed by just one power of
$M_P$, they may be accessible to the next round of
experiments with CPLEAR and/or DA$\phi$NE.
\pr
\section{Connection with Cosmology }
\pr
We complete this review with a discussion of cosmology in
the context of our non-critical approach to string theory,
commenting
on
inflation, entropy generation, variations in
fundamental parameters, and the cosmological constant \cite{emnharc}.
The first explicit cosmological string theory with a
time-dependent background was that discovered in
\cite{aben3}. It has a dilaton field that depends
linearly on the time variable $t$, and the string
black hole can be regarded as a Minkowski rotation
of this model, in which the $t$-dependence of the
dilaton field is replaced by the corrsponding
dependence on the radial coordinate $r$. It was
suggested in \cite{aben3} that the Universe might
make quantum transitions between models with different
values of the central charge associated with
the dilaton/time variable, but this suggestion
was not worked out in detail. The mechanism for
such quantum transitions via instantons
has now been worked out in the black hole case \cite{emnharc,emnshort,
emndollar}.
As discussed in section 5, it leads (\ref{thirteenv}) to a
scale-dependent value of the level parameter $k$,
corresponding in turn to a time-dependence
\be
k(t) \simeq k e^{-4\pi \beta^I T_0 t}
\label{cone}
\ee
in the dilute-gas approximation, where we recall
that the corresponding instanton $\beta$-function
\be
\beta^I = -(k/2)g^I
\label{ctwo}
\ee
in the large-$k$ limit is negative, leading to
an increase in the effective level parameter $k$,
and hence an approach to the flat limit
${k \rightarrow \infty}$. In the cosmological
context, this implies a slowing down in the rate of
expansion of the Universe. According to the analogy
with the quantum Hall effect \cite{emndua}, this process is analogous
to a series of transitions between different
conductivity plateaux.
\pr
Several questions arise in this picture, including
the following. What was the initial state of the Universe?
Is there any analogue of inflation in this approach,
particularly as regards entropy generation, which is
an essential feature of our non-critical treatment
of string theory, as seen in equation (\ref{sevent})?
Do fundamental physics parameters such as the
velocity of light $c$ and Planck's constant ${\nd h}$
vary during the cosmological expansion \cite{emnharc}?
How does the
effective cosmological constant relax to zero, as it
should do in the flat-space limit?
\pr
In partial answer to the first of these questions, we
recall that the central charge
\be
c = 2(k + 1)/(k - 2)
\label{cthree}
\ee
becomes infinite in the limit ${k \rightarrow 2}$. The
dilute gas approximation is not reliable here,
but is suggestive that the Universe started from such a
limit. In the picture of section 4, as shown in figure 2,
this limit would correspond to the infrared limit of
the renormalization group flow. In the case of the
string black hole, it
corresponds to the region close to the core
where there is an appropriate
description \cite{emntop} in terms of a twisted ${N = 2}$
supersymmetric theory, which is equivalent to a
topological field theory in which the concepts
of space and time break down.
Thus we are led
to the conclusions that the origin of the Universe
is presumably also described by such a topological
field theory, and that
the concepts of space and time also break down
at the beginning of the Universe.
\pr
Our simple cosmological scenario provides a qualitative
picture of the entropy production rate in the Universe.
In our framework,
the rate of entropy increase with
time
is given by \cite{emnqm,emnharc}
\be
\partial _t S = \beta ^i G_{ij} \beta ^j S \qquad ; \qquad
G_{ij}=2 |z|^4 <V_i V_j >
\label{entropy}
\ee
where the unitarity requirement of the world-sheet theory implies
the positivity of the Zamolodchikov metric \cite{zam} $G_{ij} >0$.
Using the C-theorem \cite{zam}, especially in
its string formulation \cite{mavc} on the fiducial-metric world-sheet,
one may write
\be
 \beta ^i G_{ij} \beta ^j =
 \partial _t C(g) \qquad :
\qquad
 C(g) =
 -\frac{1}{12}
 \int d^Dy \sqrt{G} e^{-2\Phi} <TT> + \dots
\label{ctheorem}
\ee
 In
this expression, the
$y$ denote  target spatial coordinates,
$\Phi$ is the dilaton field, and $T \equiv
T_{zz}$ is a component of the world-sheet stress tensor.
The $\dots $ denote the remaining two-point functions that appear in the
Zamolodchikov C-function \cite{zam}, which
involve the trace $\Theta $
of the stress tensor, i.e.
$<T \Theta >$ and $ <\Theta \Theta >$. Taking into account the
off-shell corollary of the C-theorem, $\frac{\delta C(g)}{\delta g^i}
=G_{ij}\beta ^j$, it can readily be shown \cite{mavr} that
such terms
can always be removed by an appropriate renormalization-scheme choice,
that is by
appropriate redefinitions of the
renormalized couplings $g^i$, and hence play no r\^ole in the physics.
Thus, one can
solve (\ref{entropy}) for the entropy $S$
in terms of the Zamolodchikov
C-function
\be
S(t)=S_0 e^{-\frac{1}{12}\int _0^t \int d^Dy \sqrt{G}e^{-2\Phi}
<TT> + \dots}
\label{entrzam}
\ee
where the minus sign in the exponent indicates the opposite
flow of the time $t$ with respect to the renormalization-group
flow.
Expression
(\ref{entrzam}) reduces a complicated target-space computation
of entropy production in an inflationary scenario to a
conformal-field-theory computation of two-point functions
involving components of the stress tensor of a first-quantized string.
We observe from (\ref{entropy}) that
the rate of entropy increase
is maximized on the
maximum-$\beta ^i$
surface in coupling constant
space.
At late stages of the inflationary era,
i.e. close to the ultraviolet fixed point, the rate of change of $S$
is strongly suppressed, due to the smallness of the
$\beta ^i$.
\pr
In order to discuss the possible variation of
fundamental physical parameters during the expansion
of the Universe \cite{emnharc}, we first recall the relation
between a string black hole mass and the level
parameter $k$:
\be
M/M_{Planck} =
\sqrt{\frac{1}{k(t)-2}}
e^{const}
\label{cfour}
\ee
It is well known that light cones are distorted
in the presence of a black hole. Specifically,
the exact space-time background metric of the
black hole Wess-Zumino model has the following
asymptotic form for large $r$:
\be
ds^2 = 2 (k(t)-2)(dr^2 -
\frac{k(t)}{k(t)-2}dt^2 )
\label{cfive}
\ee
This implies a $k$-dependence of the apparent
velocity of light, which becomes a time-dependence
as a result of equation (\ref{cone}):
\be
c_q =c\sqrt{\frac{k(t)}{k(t)-2}}
\label{csix}
\ee
where $c$ is the usual flat-space velocity. We
note that the fact that $c_q$ $\rightarrow$ $\infty$ as
$k$ $\rightarrow$ $2$, corresponding to the broadening out
of the effective light-cone,
is consistent with the suggestion
made above that the concepts of space and time break down in
this limit. Specifically, in a Robertson-Walker-Friedmann
universe the horizon distance $d$ in co-moving
coordinates over which an observer can look back is \cite{rwf}
\be
d=\int c_q (t) = \int dt \sqrt{\frac{k(t)}{k(t)-2}}
\label{horizon}
\ee
which is larger than the naive estimate ${d = c t}$.
Indeed, the horizon distance could even become
infinite if ${k(t) \rightarrow 2}$ in a suitable
way as ${t \rightarrow 0}$, but this conjecture
takes us beyond the dilute-gas approximation
where we can compute reliably.
\pr
The time-dependence of string physics is also reflected
in a computation of the string position-momentum
uncertainty relation. Defined appropriately to incorporate
curved gravitational backgrounds, this uncertainty can
be expressed as \cite{emnharc}
\be
(\Delta X \Delta P)_{min} \equiv \hbar _{eff} (t) = \hbar (1 +
O(\frac{1}{k(t)})
\label{cseven}
\ee
where ${\Delta A}  = (< A^2 > - < A >^2)^{\frac{1}{2}}$,
$ < \dots > $ denotes a
$\sigma$-model vacuum expectation value, and $\hbar$
is the critical-string Planck's constant. The
string uncertainty relation introduces a minimum
length $\lambda_s$, that in our case also decreases
with time \cite{emnharc}:
\be
\lambda _s (t) \equiv (\frac{\hbar _{eff} (t)\alpha ' (t) }
{c_q (t)^2})^{\frac{1}{2}}=\lambda _s^0 (1 + O(\frac{1}{k(t)}) )
\label{ceight}
\ee
We mention in passing that the effective string
Regge slope is also time-dependent \cite{emnharc}:
\be
\frac{\alpha ' (t)}{\alpha '^0} = \frac{c_q (t)^2}{c^2}
\label{cnine}
\ee
which stems from the relation between $k$ and
$\alpha '$  in this model. It is also
worth mentioning that this cosmological model
exhibits certain Jeans-like instabilities \cite{sanchez}
leading to the  exponential growth of
low-energy string modes at
finite $k$ \cite{emnharc}, thereby providing
a scenario for string-sustained inflation \cite{turok}.
\pr
The non-critical string  scenario for the expanding
Universe described in the preceeding paragraphs
offers the prospect of solving
the three basic
problems of the standard-model cosmology
 in a manner reminiscent
of conventional inflation \cite{guth}.
The {\it horizon problem} could be solved by the
enhanced look-back distance (\ref{horizon}), and/or
the breakdown of the normal concepts of space and
time in a transition to a topological phase close to
the infrared fixed point. The {\it flatness}
problem could be solved by an epoch of
exponential expansion, induced by a Jeans-like instability
\cite{emnharc}.
The {\it entropy problem} could be solved by
the enhanced rate of entropy production (\ref{entrzam})
at early times.
However,
the
crucial difference in our
approach is that the fundamental scalar field, usually
termed the
{\it inflaton},
is replaced by
a world-sheet field, the Liouville mode,
in our approach.
Fluctuations of this field create the renormalization
group flow of the system that leads to the generation
of propagating
matter, in the way described above and
in previous works\cite{emndua,emndollar}.
Of course, this mode is associated with the appearance of
a target space scalar, the dilaton, but the latter is part of the
metric background. This can be seen
clearly in the two-dimensional
Wess-Zumino string theory, which may be considered as a
prototype for the description of a spherically-symmetric
($s$-wave) four-dimensional
Universe \cite{emn4d}. In this model
the dilaton belongs to the graviton level-one string multiplet,
which is a non-propagating (discrete)
string mode, and as such can only
exist as a background, in contrast to a massless `tachyon' mode,
which propagates and scatters.
\pr
In our approach to the cosmological constant question, we
start by considering the following
one-loop results for the dilaton and graviton
$\beta$-functions in bosonic $\sigma$-models \cite{tseytlin,osborn}:
\bea
   \beta ^\Phi \equiv \frac{d \Phi }{d \phi}
   &=& -\frac{2}{\alpha '} \frac{\delta c}{3} + \nabla ^2 \Phi
- (\nabla \Phi )^2  \nn \\
  \beta ^G _{MN} \equiv \frac{d G_{MN}}{d\phi}
  &=&- R_{MN} - 2 \nabla _M \nabla _N \Phi
\label{betafunceq}
\eea
where $\Phi$ is the dilaton field and
$\phi$ denotes our covariant Liouville cutoff (c.f. the relative
minus sign compared
with the notation
of ref. \cite{tseytlin,osborn} where the cutoff is defined
with the dimensions of mass), and
$\delta c = C-26$, the $26$ coming from the space-time
reparametrization
ghosts.
If
the central charge of the
theory is not 26, as is the case of non-critical bosonic strings,
then
a
cosmological constant term appears
in the target space effective action.
The form of this target-space
action,
whose variations
yield the $\beta$-functions
(\ref{betafunceq}), reads:
\be
 {\cal I}=\frac{2}{\alpha '}
 \int d^Dy \sqrt{G}e^{-\Phi} \{ \frac{1}{3}\delta
 c - \alpha '
(R + 4 (\nabla \Phi )^2 + \dots ) \}
\label{effbossigma}
\ee
where the $\dots$ denote other fields in the theory that we shall not
use explicitly. We now notice that the effects of the tachyons
in our two-dimensional target-space string model amount to a shift
of the level parameter $k(\phi)$ with the
renormalization
group scale. This is the result of the combined
effects of tachyon and instanton deformations, the latter
representing higher-genus instabilities \cite{emndec,cohen,
emnharc,emndollar}.
The
instantons alone, as
irrelevant deformations,
produce an initial instability by inducing an
increase of
the central charge,
which then flows
downhill towards 26
in the presence of relevant matter (tachyon) couplings.
Hence there is
a running central charge $c(\phi) > 26$, according to
the C-theorem \cite{zam}, that will, in general, imply a non-vanishing,
time-varying (running), {\it positive}
cosmological constant, $\Lambda (\phi)$,
for the background of (\ref{cfive}). Its precise
form is determined by consistency with the equations (\ref{betafunceq}).
\pr
For simplicity,
we assume that the only effect of the dilaton
is a constant contribution to the scale anomaly, which is certainly
the case of interest. This allows one to decouple
$\Phi$ in the
field equations obtained from (\ref{effbossigma}). Then
the latter read
\bea
\frac{\delta {\cal I}}{\delta \Phi}&=&\Lambda (\phi) - R  \nn \\
\frac{\delta {\cal I}}{\delta G_{MN}}&=&-R_{MN} + \frac{1}{2}G_{MN}R
\eea
In two dimensions the second equation is satisfied identically.
Decoupling of the dilaton field also implies that the first equation
yields
\be
            R=\Lambda (\phi)
\label{curvcosmo}
\ee
The metric background (\ref{cfive}) has a maximal symmetry in its
space part. To make the analysis more general, we extend the
background to $d=2 + \epsilon$ dimensions, keeping the maximal
symmetry in the spatial part of the metric \footnote{The
$G_{00}$-component depends  at most on the time $\phi$ and can be
absorbed in a redefinition of the time variable. It will not
be of interest to us here.}. Relating time to the Liouville field
as we have discussed earlier, we find
the following solution for $\Lambda (\phi)$\footnote{We remark
that
a similar
equation
has also been
considered in ref. \cite{kogan}, but the flow  of time in that
reference coincides
with the renormalization group flow. In such a case,
one gets sensible results only for negative initial values of the
cosmological constant, contrary to our case where we have a
vanishing cosmological constant asymptotically, starting from
positive initial values.}
\be
    \Lambda (t)=\frac{\Lambda (0)}{1 + t
     \frac{\Lambda (0)}{d-1}} \qquad ; \qquad t\equiv -\phi > 0
\label{runnincosmo}
\ee
which for positive $\Lambda$ implies an
asymptotically-free
cosmological constant $\beta$-function, thereby leading to
a vanishing cosmological constant at the ultraviolet fixed point
on the world-sheet.
\pr
The rate of the decrease of $\Lambda (\phi)$
is determined by its initial value at the infrared
fixed point, where we conjectured that
the theory makes a transition to a topological (twisted
$N=2$ supersymmetric) $\sigma$-model. It is of great
interest to estimate
this value in our two-dimensional model. This
can be done by noticing that
\be
        \Lambda (0)=\frac{2}{\alpha '(0)}\frac{ c(0)-26}{3}
\label{irlam}
\ee
where $c(0)-26
 =\frac{3 k(0)}{k(0)-2} -27 \simeq \frac{3k(0)}{k(0) -2}$,
given that $k(\phi) \rightarrow 2 $ as $\phi \rightarrow 0$. Thus,
taking into account (\ref{irlam}) one observes that
$\Lambda (0)$ is determined by the critical string tension, as it
should be, given that $\alpha '_0$
is the only scale in the problem
(or equivalently the minimal  string length): the result is
\be
     \Lambda (0) = \frac{2}{\alpha '_0}
\label{critalpha}
\ee
The latter result implies a really fast decay of the cosmological
constant in this model. Notice that the finite initial
value of $\Lambda (0)$ implies from (\ref{curvcosmo})
 no curvature singularity in the
Euclidean model at the origin of target space $r=0$,
as is indeed the case of the two-dimensional
black-hole model of ref. \cite{witt}, given that
this point
is a pure coordinate singularity. The above analysis
for the cosmological constant, therefore, applies most likely
to singularity-free inflationary universes \cite{barrow}.
It is understood that until the
precise behaviour of the running couplings near the infrared fixed
point is found, there will always be uncertainties in the above
estimates. String perturbation theory is not applicable near the
topological phase transition, and the infinities we get in the various
running couplings constitute
an indication of this. In the complete theory,
these infinities should be absent.
\pr
\section{Outlook}
\pr
We have outlined in these lectures an approach to
non-equilibrium quantum statistical mechanics,
black holes, time, quantum mechanics and
cosmology that is based on non-critical string theory,
with time described by a Liouville field. Our basic
aim has been to understand some of the qualitative
features of quantum fluctuations in the structure of
space-time, and their physical consequences, i.e., to
understand foam. We have tried to short-circuit the
general ignorance of string field theory by using a
sort of mini-superspace approach, exploiting our
knowledge of one particular class of such fluctuations,
namely string black holes, and arguing that their
consequences are likely to be quite general.
Specifically, we expect that many other types of
space-time fluctuation will tend to suppress quantum
coherence in the manner discussed here for the black hole
case.
\pr
We are aware that many details of these ideas
remain to be worked out, and that many questions could be
raised. However, we believe that our approach manifests such
a high degree of internal consistency, and brings together
so many apparently unconnected features of different areas
of physics, that it is worthy of further constructive study.
\pr
There are several aspects of our work that we have not discussed
here in any detail. These include the possible r\^ole of the type
of decoherence discussed here in the transition between
microscopic quantum physics and macroscopic classical physics.
We have already given some discussion of this, and see a
connection with ideas of Penrose \cite{Pen}
that we plan to discuss
elsewhere. It also seems that our approach can put the problem
of measurement in a new light. These are just some examples
of basic physical problems where others have conjectured that
quantum gravity may play a r\^ole. In string theory, we have
for the first time a consistent quantum theory of gravity
in which these questions can be addressed in a meaningful
way. We have started to provide some answers. Some of them
may be incomplete or wrong in detail, but we believe we
have put our fingers on some important aspects of the truth.
We urge the reader
to examine our approach with a cool
head.
\pr
\noindent {\Large {\bf Acknowledgements }} \\

One of us (N.E.M.) thanks C. Ktorides for valuable discussions about
his work.
The work of N.E.M. is supported by a EC Research Fellowship,
Proposal Nr. ERB4001GT922259.
The work of D.V.N.
is partially supported by DOE grant DE-FG05-91-ER-40633.

\newpage
\pr
{\Large {\bf Figure Captions}}
\pr
\nk {\bf Figure 1 } - Contour
of integration in the analytically-continued
(regularized) version of $\Gamma (-s)$ for $ s \in Z^+$.
This is known in the literature as the Saalschutz contour,
and has been used in
conventional quantum field theory to relate dimensional
regularization to the Bogoliubov-Parasiuk-Hepp-Zimmermann
renormalization method.
\pr
\nk {\bf Figure 2 } - Schematic repesentation
of the evolution of the world-sheet area as the renormalization
group scale moves along the contour of fig. 1.
\pr
\nk {\bf Figure 3 }- The
$(\beta, \gamma)$ plane on
a logarithmic
scale for $\beta > 0$. We plot
contours of the conventional
$CP$-violating parameter $|\epsilon |$, evaluated from the
$K_L \rightarrow 2\pi $ decay rate. The dashed-double-dotted
band is that allowed
at the one-standard-deviation level by the comparison
between measurements of the $K_L \rightarrow 2\pi $ decay rate
and the $K_L$ semileptonic decay asymmetry $\delta _L $.
The dashed line delineates the boundary
of the region allowed by the
present experimental upper limit on $K_S \rightarrow 3\pi ^0$
decays $(R_{2\pi}^L-R_{3\pi}^S)$
and a solid line delineates the boundary
of the region allowed by
a recent preliminary measurement of the $K_S$
semileptonic decay asymmetry $\delta _S$.
A wavy line bounds approximately the region of $|\beta |$
which may be prohibited by intermediate-time measurements of
$K \rightarrow 2\pi$ decays.
\pr

\nk {\bf Figure 4 }-  As in Fig. 3, on a
logarithmic scale for $\beta < 0$.

\pr
\nk {\bf Figure 5 }-  As in Fig. 3, on a
linear scale for the neighborhood of $\beta=0$.


\newpage
\begin{thebibliography}{99}
\bibitem{aben3} I. Antoniadis, C. Bachas, J. Ellis
and D.V. Nanopoulos, Phys. Lett. B211 (1988), 393;
Nucl. Phys. B328 (1989), 117; Phys. Lett. B257 (1991), 278.
\bibitem{emnqm} J. Ellis, N.E. Mavromatos
and D.V. Nanopoulos, Phys. Lett. B293 (1992), 37.
\bibitem{emnshort}
J. Ellis, N.E. Mavromatos and
D.V. Nanopoulos, CERN, ENS-LAPP and Texas A \& M Univ. preprint,
CERN-TH.6896/93, ENS-LAPP-A426-93, CTP-TAMU-29/93; ACT-09/93 (1993);
hep-th/9305116.
\bibitem{emndollar}
J. Ellis, N.E. Mavromatos and
D.V. Nanopoulos, CERN, ENS-LAPP and Texas A \& M Univ. preprint,
CERN-TH.6897/93, ENS-LAPP-A427-93, CTP-TAMU-30/93; ACT-10/93 (1993);
hep-th/9305117.
\bibitem{witt} E. Witten, Phys. Rev. D44 (1991), 314.
\bibitem{rg} See, for instance,
A.M. Polyakov, {\it Gauge Fields and
Strings} (Harwood, New York 1984).
\bibitem{hawk} S. Hawking, Comm. Math. Phys. 87 (1982), 395.
\bibitem{bek} J. Bekenstein, Phys. Rev. D12 (1975), 3077.
\bibitem{hawk2} S. Hawking, Comm. Math. Phys. 43 (1975), 199.
\bibitem{ehns} J. Ellis, J.S. Hagelin, D.V. Nanopoulos and
M. Srednicki, Nucl. Phys. B241 (1984), 381.
\bibitem{misra} B. Misra, I. Prigogine and M. Courbage,
{\it Physica} A98 (1979), 1;
\par I. Prigogine, {\it Entropy, Time, and
Kinetic Description}, in {\it Order and Fluctuations
in Equilibrium and Non-Equilibrium Statistical Mechanics},
ed G. Nicolis {\it et al.} (Wiley, New York 1981);
\par B. Misra and I. Prigogine, {\it Time, Probability and Dynamics},
in {\it Long-Time Prediction in Dynamics}, ed G. W. Horton,
L. E. Reichl and A.G. Szebehely (Wiley, New York 1983).
\bibitem{misra2} B. Misra, {\it Proc. Nat. Acad. Sci. U.S.A.}
75 (1978), 1627.
\bibitem{santilli} R. M. Santilli, Hadronic
J. 1 (1978), 223 , 574 and 1279 ;
 {\it Foundations of Theoretical Mechanics}, Vol.  I
(1978) and II (1983)  (Springer-Verlag, Heidelberg-New York);
\par For an application of this approach to
dissipative statistical systems, which is directly
relevant to our work here, see:
J. Fronteau, A. Tellez-Arenas
and R.M. Santilli, Hadronic J. 3 (1979), 130;
\par J. Fronteau, Hadronic J. 4 (1981), 742.
\bibitem{Ktorides} J.P. Constantopoulos and
C.N. Ktorides, J. Phys. A17 (1984), L29.
\bibitem{emncpt} J. Ellis, N.E. Mavromatos and
D.V. Nanopoulos, Phys. Lett. B293 (1992), 142;
CERN and Texas A \& M Univ. preprint
CERN-TH. 6755/92; ACT-24/92;CTP-TAMU-83/92; hep-th/9212057.
\bibitem{emnharc} J. Ellis, N.E. Mavromatos
and D.V. Nanopoulos, preprint CERN-TH.7000/93, CTP-TAMU 66/93,
ENSLAPP-A-445/93, OUTP-93-26P, hep-th/9311148,
{\it Opening lecture
at HARC workshop on ``Recent Advances in the
Superworld", The Woodlands, Texas (USA), April 14-16 1993}, to
appear in the proceedings.
\bibitem{emndua} J. Ellis, N.E. Mavromatos and D.V. Nanopoulos,
Phys. Lett. B289 (1992), 25; {\it ibid} B296 (1992), 40.
\bibitem{yung} A.V. Yung,
Int. J. Mod. Phys. A9 (1994), 591.
\bibitem{wald} R. Wald, Phys. Rev. D21 (1980), 2742.
\bibitem{vernon} R.P. Feynman and F.L. Vernon Jr., Ann. Phys.
(NY) 94 (1963), 118.
\bibitem{cald} A.O. Caldeira and A.J. Leggett, Ann. Phys.
149 (1983), 374.
\bibitem{davydov} Y.R. Shen, Phys. Rev. 155 (1967), 921;
\par A.S. Davydov and A. A. Serikov, Phys. Stat. Sol.
B51 (1972), 57;
\par B.Ya. Zel'dovich, A.M. Perelomov, and V.S. Popov,
Sov. Phys. JETP 28 (1969), 308;
\par For a recent review see : V. Gorini {\it et al.},
Rep. Math. Phys. Vol. 13 (1978), 149.
\bibitem{emntop} J. Ellis, N.E. Mavromatos and
D.V. Nanopoulos, Phys. Lett. B288 (1992), 23.
\bibitem{stringbook} M.B. Green, J.H. Schwarz and E. Witten,
{\it String Theory}, Vol. I and II (Cambridge Univ. Press 1986).
\bibitem{ginsp} See, for instance, P. Ginsparg, Nucl. Phys.
B295 (1988), 153.
\bibitem{DDK}F. David, Mod. Phys. Lett. A3 (1988), 1651;
\par J. Distler and H. Kawai, Nucl. Phys. B321 (1989), 509.
\bibitem{mm} N.E. Mavromatos and J.L. Miramontes,
Mod. Phys. Lett. A4 (1989), 1847.
\bibitem{shore} G. Shore, Nucl. Phys. B286 (1987), 349.
\bibitem{osborn} H. Osborn, Nucl. Phys.
B294 (1987), 595; {\it ibid}
B308 (1988), 629; Phys. Lett. B222 (1989), 97.
\bibitem{aben} J. Polchinski, Nucl. Phys. B324 (1989), 123;
\par D.V. Nanopoulos, in {\it Proc. Int. School
of Astroparticle Physics}, HARC-Houston (World Scientific, Singapore,
1991),
p. 183.
\bibitem{zam} A.B. Zamolodchikov, JETP Lett. 43 (1986), 730;
Sov. J. Nucl. Phys. 46 (1987), 1090.
\bibitem{santos} N.E. Mavromatos, J.L. Miramontes and
J.M. Sanchez de Santos, Phys. Rev. D40 (1989), 535.
\bibitem{vafa} C. Vafa, Phys. Lett. B212 (1987), 27.
\bibitem{emnnew} J. Ellis, N.E. Mavromatos and D.V. Nanopoulos,
to appear.
\bibitem{chaudh} S. Chaudhuri and J. Lykken, Nucl. Phys B396 (1993),
270.
\bibitem{emn1}
J. Ellis, N.E.  Mavromatos and
D.V. Nanopoulos,
Phys. Lett. B267 (1991), 465; {\it ibid}
B272 (1991), 261.
\bibitem{matrix} E. Br\'ezin and V.A. Kazakov,
Phys. Lett. B236 (1990), 144;
\par M. Douglas and A. Shenker, Nucl. Phys.
B335 (1990), 635;
\par D. Gross and A.A. Migdal, Phys. Rev. Lett.
64 (1990), 127;
\par For a recent
review see, e.g., I. Klebanov,
in {\it String Theory and Quantum Gravity}, Proc. Trieste
Spring School 1991, ed. by J. Harvey et al.
(World Scientific, Singapore, 1991), and references therein.
\bibitem{emnwhair} J. Ellis, N.E. Mavromatos and
D.V. Nanopoulos, Phys. Lett. B284 (1992), 43.
\bibitem{callan} V.A. Rubakov, Nucl. Phys. B203 (1982), 311;
\par C. Callan, Nucl. Phys. B212 (1983), 391.
\bibitem{Li} M. Goulian and M. Li, Phys. Rev. Lett.
66 (1991), 2051.
\bibitem{bershadsky} M. Bershadsky and D. Kutasov, Phys. Lett.
B266 (1991), 345.
\bibitem{kogan2} I. Kogan, Phys. Lett. B265 (1991), 269.
\bibitem{BPZ} T. Roy and A. Roy Chowdhuri,
Phys. Rev. D15 (1977), 3768.
\bibitem{eguchi} T. Eguchi, Mod. Phys. Lett. A7 (1992), 85.
\bibitem{coleman} S. Coleman, Phys. Rev. D15 (1977), 2929;
1248 (E) (1977).
\par C. Callan and S. Coleman, Phys. Rev. D16 91977), 1762.
\bibitem{ovrut} B.A. Ovrut and S. Thomas, Phys. Rev. D43 (1991),
1314.
\bibitem{kutasov} D. Kutasov, Mod. Phys. Lett.
A7 (1992), 2943.
\bibitem{pruisk} A. Pruisken, Nucl. Phys. B290[FS20] (1987),
61.
\bibitem{yung3} A.V. Yung, Swansea preprint SWAT 94/22
(1994).
\bibitem{emndec} J. Ellis, N.E. Mavromatos and
D.V. Nanopoulos, Phys. Lett. B276 (1992), 56.
\bibitem{fischler} W. Fischler and L. Susskind, Phys.
Lett. B171 (1986), 262; {\it ibid} B171 (1986), 383.
\bibitem{cohen} E. Cohen, H. Kluberg-Stern,
and R. Peschanski, Nucl. Phys. B328 (499) (1989);
\par E. Cohen, H. Kluberg-Stern, H. Navelet,
and R. Peschanski, Nucl. Phys. B347 (802) (1990).
\bibitem{supra} I. Affleck, M. Dine and N. Seiberg,
Nucl. Phys. B241 (1984), 493.
\bibitem{witt2} E. Witten, Nucl. Phys. B373 (1992), 187.
\bibitem{mukhi} S. Mukhi and C. Vafa, Harvard Univ. and
Tata Inst. preprint HUTP-93/A002; TIFR/TH/93-01 (1993).
\bibitem{mueller} A. Mueller, Phys. Rep. 73 (1981), 237.
\bibitem{khoze} V.V. Khoze and A. Ringwald, Nucl.
Phys. B355 (1991), 351.
\bibitem{yung1} I.I. Balitsky and A.V. Yung, Phys. Lett.
B168 (1986), 113;
\par A.V. Yung, Nucl. Phys. B297 (1988), 47.
\bibitem{dorinst} N. Dorey, Los Alamos National Lab. preprint,
LA-UR-92-1380 (1992), Phys. Rev. D to be published.
\bibitem{BW} J. Bossart and Ch. Wiesendanger,
ETH and Univ. of Z\"urich preprint, ETH-TH/91-42;
ZH-TH-32/91 (1991).
\bibitem{libby} A. Pruisken, Nucl. Phys. B235[FS11] (1984), 277;
\par H. Levine, S. Libby and
A. Pruisken, Phys. Rev. Lett. 51 (1983), 1915;
Nucl. Phys. B240[FS12] (1984),
30;49;71.
\bibitem{lee} Kimyeong Lee, Nucl. Phys. B373 (1992), 735.
\bibitem{lutken} C.A. L\"utken and G.G. Ross,
Oxford preprint OUTP-92-22P (1992), and references therein.
\bibitem{emohn} J. Ellis, S. Mohanty and D.V. Nanopoulos,
Phys. Lett. B221 (1989), 113.
\bibitem{eliashvilli} M Eliashvili,
LAPP (Annecy) preprint ENSLAPP-A-462/94  (1994).
\bibitem{laughlin} R.B. Laughlin, in {\it The Quantum
Hall Effect}, ed. R.A. Prange and S.M. Girvin (Springer-Verlag
New York 1990).
\bibitem{trug} A. Capelli, C. Trugenberger and G. Zemba,
Nucl. Phys. B396 (1993), 465 ;
\par S. Iso, D. Karabali and B. Sakita, Phys. Lett. B296
(1992), 143.
\bibitem{bakas} I. Bakas and E. Kiritsis,
Int . J. Mod. Phys. A7 (Suppl. 1A) (1992), 55.
\bibitem{verl} R. Dijkgraaf, E. Verlinde and H. Verlinde,
Nucl. Phys. B371 (1992), 269.
\bibitem{bakir} I. Bakas and E. Kiritsis, ENS (Paris) preprint
LPTENS-92-30 (1992).
\bibitem{cpt} G. Luders, Ann. Phys. (N.Y.) 2, (1957), 1.
\bibitem{cptwitt} E. Witten, Comm. Math. Phys. 109 (1987), 525.
\bibitem{cptstring} V. A. Kosteleck\'y  and R. Potting, Nucl. Phys.
B359 (1991), 545.
\bibitem{sonoda} H. Sonoda, Nucl. Phys. B326 (1989), 135.
\bibitem{elmn} J. Ellis, J. Lopez, N.E. Mavromatos
and D.V. Nanopoulos, in preparation ;
\par J. Lopez, preprint
CPT-TAMU 38/93 (1993), talk
at {\it HARC workshop on ``Recent Advances in the
Superworld", The Woodlands, Texas (USA), April 14-16 1993}, to
appear in the proceedings.
\bibitem{hp} P. Huet and M. Peskin, SLAC (Stanford) preprint
SLAC-PUB-6454 (1994), hep-ph/9403257.
\bibitem{pdg} {\it Review of Particle Properties},
Particle Data Group, Phys. Rev. D45 (No 11, part II) (1992), 1.
\bibitem{mavc} N.E. Mavromatos and J.L. Miramontes,
Phys. Lett. B212 (1988), 33;
\par N.E. Mavromatos, Phys. Rev. D39 (1989), 1659.
\bibitem{mavr} N.E. Mavromatos, Mod. Phys. Lett. A3 (1988), 1079.
\bibitem{rwf} See, for instance, S. Weinberg {\it Gravitation and
Cosmology} (Wiley, New York 1972 ).
\bibitem{sanchez} N. Sanchez and G. Veneziano, Nucl. Phys.
B333 (1990), 253.
\bibitem{turok} N. Turok, Phys. Rev. Lett. 60 (1988), 549.
\bibitem{guth} See, for instance, A.H. Guth, {\it Phase Transitions
in the Embryo Universe}, Phil. Trans. Roy. Soc. Lond. A307 (1982),
141.
\bibitem{emn4d} J. Ellis, N. Mavromatos and D.V. Nanopoulos,
Phys. Lett. B278 (1992), 246.
\bibitem{tseytlin} A.A. Tseytlin, Phys. Lett. B178 (1986), 34.
\bibitem{kogan}I. Kogan, Univ. of British Columbia preprint,
UBCTP 91-13 (1991).
\bibitem{barrow} I. Antoniadis, G.F.R. Ellis, J. Ellis,
C. Kounnas and D.V. Nanopoulos, Phys. Lett. B191 (1987), 393;
\par J. Barrow and N. Deruelle, Nucl. Phys. B297 (1988), 733.
\bibitem{Pen} R. Penrose, {\it The Emperor's New Mind}
(Oxford Univ. Press, 1989).
\end{thebibliography}
\end{document}